\documentclass[12pt, draftclsnofoot, onecolumn]{IEEEtran}
\usepackage{geometry}
\usepackage{amsmath,amssymb}
\usepackage[dvips]{graphicx}
\usepackage{amsfonts}
\usepackage[mathscr]{eucal}
\usepackage{subfigure}
\usepackage{float}
\usepackage{latexsym}
\usepackage{amsthm}
\usepackage{exscale}
\usepackage[mathscr]{eucal}
\usepackage{bm}
\usepackage[dvipsnames]{color}
\usepackage{cases}
\usepackage{color}
\usepackage{epsfig}
\usepackage{stfloats}
\usepackage[ruled]{algorithm2e}
\usepackage[verbose,nospace,sort]{cite}
\usepackage{tabularx,ragged2e}
\usepackage{multirow}
\usepackage{multicol}
\usepackage{balance}
\usepackage{caption}
\usepackage{setspace}
\usepackage{multirow}
\usepackage{multicol,afterpage,wrapfig}
\usepackage{pifont}
\usepackage{array}

\theoremstyle{definition}

\usepackage{algpseudocode}

\graphicspath{{./figs/}}

\newtheorem{lemma}{Lemma}

\newtheorem{Prop}{Proposition}
\newtheorem{Rem}{Remark}

\newenvironment{myproof} {{\it{Proof:}}}{\hfill$\square$}

\begin{document}

\title{Anchor-Assisted Channel Estimation for Intelligent Reflecting Surface Aided Multiuser Communication}
\vspace{-0cm}
\author{
	Xinrong Guan, Qingqing Wu, \IEEEmembership{Member, IEEE,}	
	and Rui Zhang, \IEEEmembership{Fellow, IEEE}
	\thanks{
		X. Guan is with the College of Communications Engineering, Army Engineering University of PLA, Nanjing, 210007, China (e-mail: geniusg2017@gmail.com). 
		Q. Wu is with the State Key Laboratory of Internet of Things for Smart City, University of Macau, Macau, China 999078 (email: qingqingwu@um.edu.mo).
		R. Zhang is with the Department of Electrical and Computer Engineering, National University of Singapore, 117583, Singapore (e-mail: elezhang@nus.edu.sg). 	This work was supported in part by the National Natural Science Foundation of China under Grant 61501512 and 61671474, in part by the International Postdoctoral Exchange Fellowship Program of China under Grant 20180002, and in part by National University of Singapore under Research Grant R-261-518-005-720. Part of this work has been presented in \cite{guanCE}. 
		}
}

\maketitle
\vspace{-1.8cm}

\begin{abstract} \vspace{-2mm}
	Channel estimation is a practical challenge for intelligent reflecting surface (IRS) aided wireless communication. As the number of IRS reflecting elements or IRS-aided users increases, the channel training overhead becomes excessively high, which results in long delay and low throughput in data transmission. To tackle this challenge, we propose in this paper a new anchor-assisted channel estimation approach, where two anchor nodes, namely A1 and A2, are deployed near the IRS for facilitating its aided base station (BS) in acquiring the cascaded BS-IRS-user channels required for data transmission. Specifically, in the first scheme, the partial channel state information (CSI) on the element-wise channel gain square of the common BS-IRS link for all users is first obtained at the BS via the anchor-assisted training and feedback. Then, by leveraging such partial CSI, the cascaded BS-IRS-user channels are efficiently resolved at the BS with additional training by the users. While in the second scheme, the BS-IRS-A1 and A1-IRS-A2 channels are first estimated via the training by A1. Then, with additional training by A2, all users estimate their individual cascaded A2-IRS-user channels simultaneously. Based on the CSI fed back from A2 and all users, the BS resolves the cascaded BS-IRS-user channels efficiently. In both schemes, the channels among the fixed BS, IRS, and two anchors are estimated in a large timescale, which greatly reduces the real-time training overhead. Simulation results demonstrate that our proposed anchor-assisted channel estimation schemes achieve superior performance as compared to existing IRS channel estimation schemes, under various practical setups. In addition, the first proposed scheme outperforms the second one when the number of antennas at the BS is sufficiently large, and vice versa.
\end{abstract}

\section{Introduction}
\vspace{-0mm}
To meet the ever increasing demand for higher data rates and tremendous growth in the number of communication devices, a variety of wireless technologies have been proposed, such as ultra-dense network (UDN), massive multiple-input multiple-output (MIMO), and millimeter wave (mmWave) communication \cite{5G_1,5G_2,6G}. However, the implementation of these technologies still faces critical challenges in practice, such as excessive energy consumption, high hardware cost, and complex signal processing  \cite{Zhang}, \cite{Wu}. On the other hand, traditional approaches for combating the wireless channel fading mainly compensate for or adapt to it \cite{WC1}, \cite{WC2}, but have limited control over it. Motivated by the above, intelligent reflecting surface (IRS) has been proposed recently as a promising technology to achieve cost-effective and yet highly spectral and energy efficient wireless communication systems, by dynamically reconfiguring the wireless propagation environment  \cite{QQ, pan1, QQ_1, zhou}. Specifically, IRS is a planar array composed of a large number of low-cost, passive, and tunable reflecting elements. By varying the reflection coefficients of all elements jointly, the signal reflected by IRS can be added constructively/destructively with those from other propagation paths to enhance/suppress the received signal at designated location(s), thus achieving high passive beamforming and/or interference suppression gains. Moreover, since IRS does not require any active radio frequency (RF) chains to transmit/receive but only reflects signals, it is of much lower hardware/energy cost as compared to traditional active relays. Owing to the above appealing advantages, IRS has been recently considered as one promising technology for the future sixth-generation (6G) wireless network  \cite{whitepaper} and extensively studied under various wireless system setups, such as non-orthogonal multiple access (NOMA) \cite{zheng1}, \cite{ding}, cognitive radio \cite{guanCR}, \cite{Yu}, simultaneous wireless information and power transfer (SWIPT) \cite{QQ_2,pan,QQ_3, QQ_4}, secrecy communications \cite{Cui,guan,shen}, and so on. A comprehensive tutorial on IRS-aided wireless communications can be found in \cite{QQ_Tutorial}. 

To reap the maximum performance gains of IRS, accurate channel state information (CSI) is indispensable in practice \cite{QQ}, \cite{QQ_Tutorial}. However, channel estimation is a fundamentally challenging task in implementing  IRS-aided wireless systems due to the following reasons. First, for passive IRS that can only reflect signals without the functionalities  of signal transmission/reception, the conventional pilot-based training methods for separately estimating the base station (BS)-IRS and IRS-user channels become infeasible. As such, an alternative approach in practice is to estimate the cascaded BS-IRS-user channels based on the training signals from the BS/users by properly designing the IRS reflection pattern over time \cite{yangyf}, \cite{Jensen}. Second, IRS usually consists of a large number of reflecting elements, thus the pilot/training overhead required for estimating their cascaded channels is prohibitive in practice. To reduce the complexity of channel estimation and reflection design with a large number of IRS elements, a novel element-grouping strategy was proposed in \cite{zheng}, \cite{you}. Specifically, by grouping adjacent elements of the IRS with high channel correlation into a sub-surface with a common reflection coefficient, only the cascaded user-IRS-AP channel associated with each sub-surface needs to be estimated \cite{zheng}, \cite{you}. Moreover, the size of each sub-surface can be adjusted to achieve a flexible trade-off between training overhead reduction and passive beamforming gain \cite{zheng}, \cite{you}.

However, the above works mainly focus on channel estimation for the IRS-aided single-user system, and cannot be straightforwardly extended to the IRS-aided multiuser system. The reason is that as the number of users increases, the required pilot overhead of user-by-user successive channel estimation by applying the methods in \cite{yangyf,Jensen,zheng,you} will be proportional to the number of IRS reflecting elements (or that of sub-surfaces) times that of users, which becomes prohibitively high when the number of IRS-aided users is large. To improve the efficiency of channel estimation in IRS-aided multiuser communication systems, some initial studies have been conducted in the literature (see, e.g.,  \cite{Yuan,Chen, Liu, zheng_2,Dai}). In \cite{Yuan}, by leveraging the slow-varying BS-IRS channel and its sparsity, a new message-passing based algorithm was proposed to estimate the cascaded IRS channels efficiently. In \cite{Chen}, the cascaded IRS channel estimation was formulated as a sparse channel matrix recovery problem, for which the compressive sensing (CS) based solution was derived to achieve robust channel estimation with low training  overhead. A key observation made in \cite{Liu}, \cite{zheng_2} is that the cascaded IRS channels of all users share a common BS-IRS channel, which was exploited to devise new channel estimation schemes with significantly reduced training time for multiuser multi-antenna and orthogonal frequency division multiplexing (OFDM) systems, respectively. Specifically, the cascaded channel of a reference user is first estimated by applying the method in \cite{zheng}. Then, based on the estimated CSI of this reference user, the cascaded channels of all other users can be estimated efficiently with reduced pilot overhead, by exploiting the fact that they are scaled versions of that of the reference user. However, the above methods need to update the reference user's channel once its CSI changes, despite that the BS-IRS common channel is usually static in practice due to their fixed locations. Moreover, in \cite{Dai}, assuming a full-duplex (FD) BS, the authors proposed a dual-link (BS-IRS-BS) channel estimation scheme based on the signal sent by the BS and reflected back by the IRS. Then, by exploiting the BS-IRS-BS channel, the common BS-IRS channel is resolved for estimating the IRS-user channels and consequently the cascaded BS-IRS-user channels efficiently. However, this method requires self-interference cancellation at the BS and also suffers low received signal power due to the double path loss (to/from the IRS), especially when the distance between the BS and IRS is long.

\begin{figure}[t]
	\centering
	\includegraphics[width=3.7in]{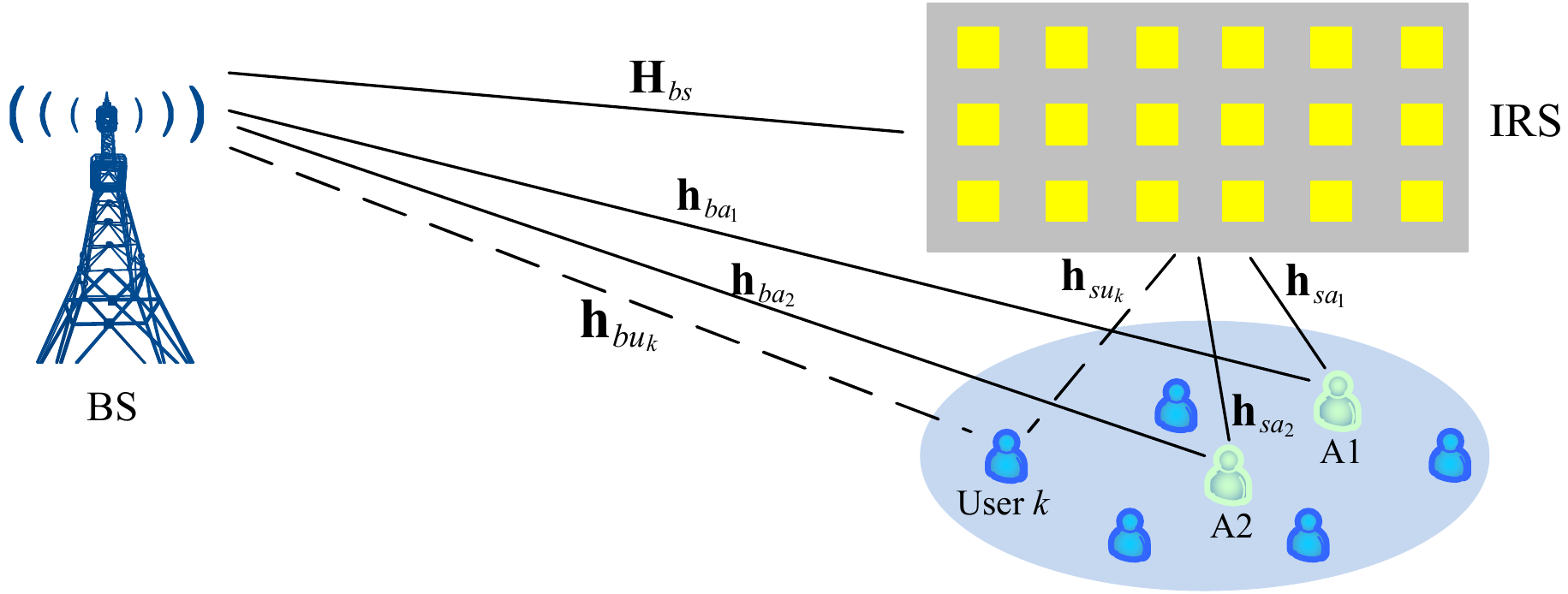}
	\caption{An IRS-aided multiuser communication system, with two anchor nodes A1 and A2 deployed near the IRS to assist in the cascaded BS-IRS-user channel estimation. }
	\label{system_model}
	\vspace{-8mm}
\end{figure}

To overcome the aforementioned limitations in the existing works, we propose in this paper a new anchor-assisted channel estimation approach for IRS-aided multiuser communication systems. As shown in Fig. 1, two anchor nodes, namely A1 and A2, are deployed near the IRS to assist in the cascaded BS-IRS-user channel estimation for its served users with their associated BS. By considering a fully passive IRS, we assume that channel reciprocity holds for all communication links and a time-division duplexing (TDD) protocol is adopted for uplink and downlink transmissions. Specifically, in the first proposed scheme, via the anchor-assisted training and feedback, the partial CSI (in terms of the element-wise channel gain square) of the BS-IRS link is first obtained at the BS. Then, by leveraging such partial CSI, the cascaded BS-IRS-user channels are efficiently resolved at the BS with additional training by the users. While in the second proposed scheme, the BS-IRS-A1 and A1-IRS-A2 channels are first estimated via the training by A1. Then, with additional training by A2, all users estimate their individual cascaded A2-IRS-user channels simultaneously. Based on the CSI fed back from A2 and users, the BS finally resolves the BS-IRS-user channels efficiently. In both schemes, we assume that the channels among the BS, IRS, and two anchors vary much more slowly than the user-involved channels since the users may change their locations while the locations of the BS, IRS and anchors are fixed. As such, the  CSI between the BS, IRS, and anchors can be estimated in a relatively much larger time scale, which thus reduces the real-time training overhead for estimating the desired cascaded BS-IRS-user channels significantly. 

The main contributions of this paper are summarized as follows:
\begin{itemize}
\item The first proposed anchor-assisted cascaded channel estimation scheme exploits the partial CSI (i.e., IRS
element-wise channel gain square) of the common BS-IRS channel for all users to reduce the real-time training overhead for estimating the IRS-user channels. We show that despite using only the square of each IRS element's channel with the BS estimated, the resultant +/- sign uncertainty on each channel does not affect the unique recovery of the cascaded BS-IRS-user channels at the BS based on the received pilot signals from the users. 
\item For the second proposed anchor-assisted scheme, the  BS-IRS-A1 and A1-IRS-A2 channels are estimated first, based on which the cascaded BS-IRS-user channels are then resolved at the BS with low training overhead. In particular, different from the first scheme, the pilot signals are sent by the anchor node A2 (instead of the users) and the users feed back their individually estimated channels (with A2 and reflected by IRS) to the BS for it to resolve the cascaded BS-IRS-user channels.   
\item We compare the above two proposed schemes in terms of training overhead, feedback complexity, and normalized mean-squared error (NMSE) performance for cascaded IRS channel estimation. In general, the first scheme requires less training time as it fully exploits the multi-antenna BS for efficient cascaded channel estimation, while the second scheme achieves higher channel estimation accuracy by exploiting the short-distance channels from A1/users to A2 via IRS (as compared to those with the BS via IRS in the first scheme). Moreover, simulation results show that both proposed schemes can achieve better NMSE performance with comparable or even significantly reduced training overhead as compared to the existing IRS channel estimation schemes \cite{Liu},  \cite{Dai}.
\end{itemize}

Note that as compared to its previous conference version \cite{guanCE}, this paper proposes a new scheme (i.e., Scheme 2) and provides a detailed comparison of the proposed schemes and the benchmark ones to demonstrate the advantages of proposed schemes.  The rest of this paper is organized as follows. Section II
presents the system model for the anchor-assisted channel estimation schemes. In Sections III and IV, we propose two anchor-assisted channel estimation schemes, respectively. Section V provides simulation results to evaluate the performance of the proposed schemes. Finally, this paper is concluded in Section VI. 

\textit{Notations:} ${{\mathbb{C}}^{M \times N}}$ denotes the space of $M \times N$ complex-valued matrices. ${\bf I}$, $\bf 1$ and $\bf 0$ denote an identity matrix, an all-one vector/matrix and an all-zero vector/matrix with appropriate dimensions, respectively. The distribution of a circularly symmetric complex Gaussian (CSCG) random vector with mean $\bf 0$ and covariance matrix $\bf \Sigma$ is denoted by ${\mathcal{CN}} \left( {\bf 0,\bf \Sigma} \right)$; and $\sim$ stands for ``distributed as''. For arbitrary matrix $\bf M$, ${\bf M}^H$ and ${\bf M}^T$ denote its conjugate transpose and transpose, respectively; rank($\bf M$), $||{\bf M}||_F$ and ${\bf M}(m,n)$ denote the rank, Frobenius norm and $(m,n)$-th element of $\bf M$, respectively. For a square matrix $\bf S$, ${\bf S}^{-1}$ denotes its inverse (if $\bf S$ is of full-rank). For a complex-valued vector $\bf x$, $\text{diag}\left( {\bf x} \right)$ denotes a diagonal matrix with each diagonal element being the corresponding element in $\bf x$. For a complex number $x$, $|x|$ and $\angle x$ denotes its modulus and angle, respectively. $\odot$ and $ \otimes $ denote the Hadamard product and the Kronecker product, respectively. Moreover, $\left\lceil \cdot \right\rceil$ denotes the ceiling operation, $\mathbb{E}(\cdot)$ denotes the statistical expectation, and $\text{mod}(x,y)$ denotes the modulus of $x$ modulo $y$. 

\vspace{-0mm}
\section{System Model}\vspace{1mm}
As shown in Fig. 1, we consider an IRS-aided multiuser communication system, which consists of a multi-antenna BS, an IRS, and a cluster of $K$ single-antenna users. The number of antennas at the BS and that of reflecting elements at the IRS are denoted by $M$ and $N$, respectively. The IRS is assumed to be deployed in the vicinity of the cluster of $K$ users due to its passive signal reflection. The channels from the BS to IRS and user $k$, $k=1,...,K$, are denoted by ${\bf{H}}_{bs} \in{\mathbb{C}^{M \times N}}$ and ${\bf{h}}_{bu_{k}}\in{\mathbb{C}^{M \times 1}}$, respectively, while that from the IRS to user $k$ is denoted by ${\bf{h}}_{su_{k}}\in{\mathbb{C}^{N \times 1}}$. We assume that two single-antenna anchor nodes A1 and A2 are deployed near the IRS to assist in its channel estimation.{\footnote{Our model applies also to the case with one anchor node with two or more antennas that can transmit and receive concurrently.}} In practice, anchor nodes can be  e.g., the controllers of the IRS and its nearby IRSs, which can communicate with the BS. The channels from the BS to A1 and A2 are denoted by ${\bf{h}}_{ba_1}\!\in\!{\mathbb{C}^{M \times 1}}$ and ${\bf{h}}_{ba_2}\!\in\!{\mathbb{C}^{M \times 1}}$, respectively, those from the IRS to A1 and A2 are denoted by ${\bf{h}}_{sa_1}\!\in\!{\mathbb{C}^{N \times 1}}$ and ${\bf{h}}_{sa_2}\!\in\!{\mathbb{C}^{N \times 1}}$, respectively, and that from A1 to A2 is denoted by ${{h}}_{a_1 a_2}$. Since IRS is a passive reflecting device, we assume that the channel reciprocity holds for each communication link between the IRS and any other node in the network; thus, we only need to acquire either the uplink or downlink channel coefficients for each link of interest. Furthermore, the channel coherence time for the links among the BS, IRS and anchors and that for the links associated with the users are denoted by $T_c$ and $T_u$, respectively (normalized by the symbol duration). Since the locations of the BS, IRS and anchors are fixed, $T_c$ is mainly affected by the scattering geometry variation, while $T_u$ is mainly affected by the user mobility. Moreover, the BS is usually deployed at a higher altitude than the users, while the IRS and anchors can also be properly deployed to minimize the effect of varying scattering geometry and thus make the BS-IRS and IRS-A1/A2 channels more stable. As a result, we assume that $T_c \gg T_u$ in our considered system. 

Let  $\!{\bf{v}}=[v_{1},...,v_{N}]^T=[\beta_{1}e^{j\theta_{1}},...,\beta_{N}e^{j\theta_{N}}]\!$ denote the equivalent reflection coefficients of the IRS, where $\beta_{n} \in \{0,1\}$ and $\theta_{n} \in [0,2\pi)$ are the reflection amplitude and phase shift of the $n$-th IRS element, respectively, $n\!=\!1,...,N$.{\footnote{More generally, we may have $\beta_{n} \in [0,1]$, which, however, requires higher hardware cost. For low-cost implementation, we consider the binary on/off control of the reflection amplitude in this paper.}} Thus, the effective channel from the BS to user $k$ is modeled as a concatenation of three components, namely, the BS-IRS link, IRS reflection, and IRS-user $k$ link, which is given by ${\bf H}_{bs}{\mathbf{\Phi }} {\bf h}_{su_k}$, where ${\mathbf{\Phi }} = \text{diag}\left({\bf{v}} \right)$. Due to the lack of signal transmission/reception capability, it is challenging to estimate ${\bf H}_{bs}$ and ${\bf h}_{su_k}$ separately at the IRS. However, from the perspective of jointly designing the active and passive beamforming at the BS and IRS for data transmission, the CSI on their cascaded channels is sufficient \cite{QQ_Tutorial}. As such, let ${\bf H}_{bsu_k}={\bf H}_{bs}\text{diag}({\bf h}_{su_k})$ denote the cascaded BS-IRS-user $k$ channel (without considering the effect of IRS reflection yet), while ${\bf H}_{bsa_1}\!=\!{\bf H}_{bs}\text{diag}({\bf h}_{sa_1})$, ${\bf H}_{bsa_2}\!=\!{\bf H}_{bs}\text{diag}({\bf h}_{sa_2})$, ${\bf h}_{a_1sa_2}\!\!=\!\!{\bf h}_{sa_2}^T\text{diag}({\bf h}_{sa_1})$ and ${\bf h}_{a_2su_k}\!\!=\!\!{\bf h}_{su_k}^T\text{diag}({\bf h}_{sa_2})$ denote the cascaded BS-IRS-A1, BS-IRS-A2, A1-IRS-A2 and A2-IRS-user $k$ channels, respectively. Note that the cascaded A-IRS-B channel equals to the B-IRS-A channel, where $\!\{\text{A, B}\} \in \{\text{BS, A1, A2, user}\}\!$, due to the channel reciprocity.

In this paper, we mainly aim to estimate the cascaded BS-IRS-user channels, i.e., ${\bf H}_{bsu_k}$, $\forall k$, as the direct channels between the BS and users, i.e., ${\bf h}_{bu_k}$'s, can be easily estimated by conventional methods based on the pilot signals sent by the users and with the IRS switched off (i.e., $v_n=0$, $\forall n$). Note that the training overhead for estimating ${\bf H}_{bsu_k}$'s would become excessively high as the number of users and/or IRS reflecting elements/sub-surfaces increases, which renders long delay and thus less time for data transmission given the limited coherence intervals of user channels. To improve the efficiency of the cascaded channel estimation, we propose two anchor-assisted schemes, which are detailed in the following two sections, respectively.

\vspace{-0mm}
\section{Exploiting Partial CSI of the Common BS-IRS Link}\vspace{1mm} 
In this section, we propose the first anchor-assisted channel estimation scheme (termed Scheme 1), which is based on the following key observation. All the cascaded BS-IRS-user channels (${\bf H}_{bsu_k}$'s) share the common static BS-IRS channel, i.e., ${\bf H}_{bs}$. This thus motivates us to decouple the estimation of each ${\bf H}_{bsu_k}$ into estimating ${\bf H}_{bs}$ and the corresponding ${\bf h}_{su_k}$, separately. Specifically, ${\bf H}_{bs}$ is (partially) estimated first with the aid of the two anchors, and then ${\bf h}_{su_k}$ is estimated with reduced training overhead based on the estimated ${\bf H}_{bs}$ and by exploiting the multiple antennas at the BS.

The channel estimation and data transmission protocol for  Scheme 1 is shown in Fig. 2. In particular, for each channel coherence time $T_{c}$, we first estimate the (partial) BS-IRS channel based on the anchor-assisted training and feedback (termed channel estimation Phase I). Then, each channel coherence time $T_{u}$ is divided into a channel estimation phase (termed channel estimation Phase II) and a subsequent data transmission phase. Specifically, in  Phase II, based on the pilot signals from the users, the BS-user channels are first estimated, and then the IRS-user channels are estimated by exploiting the estimated BS-IRS channel; finally, the cascaded BS-IRS-user channels are resolved. The details of the two channel estimation phases are described as follows.

\begin{figure*}[t]
	\centering
	\includegraphics[width=6.5in]{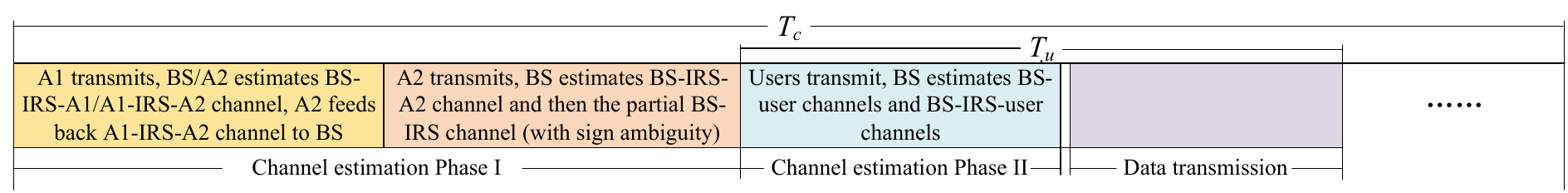}
	\caption{Channel estimation and data transmission protocol of the proposed anchor-assisted channel estimation Scheme 1.}\vspace{-4mm}
\end{figure*}

\vspace{-3mm}
\subsection{Phase I: Estimation of ${\bf{H}}_{bs}\odot{\bf{H}}_{bs}$}\vspace{-0mm}
In channel estimation Phase I, the two anchors transmit pilot symbols consecutively for estimating ${\bf{H}}_{bsa_1}$, ${\bf{H}}_{bsa_2}$ and ${\bf{h}}_{a_1sa_2}$, respectively, based on which  ${\bf{H}}_{bs}\odot{\bf{H}}_{bs}$ is estimated at the BS. We assume that in total $\tau_1+\tau_2$ pilot symbols are used in this phase. 

First, A1 transmits $\tau_1$ pilot symbols while the BS and A2 estimate ${\bf{H}}_{bsa_1}$ and ${\bf{h}}_{a_1sa_2}$, respectively. By denoting $a_{1,i_1}$ as the $i_1$-th pilot symbol transmitted by A1, $i_1\!=\!1,...,\tau_1$, the received signals at the BS and A2 are respectively given by\vspace{-0mm}
\begin{equation}
{\bf{y}}_b^{(i_1)}=\sqrt{p_1}\left({\bf{h}}_{ba_1}+{\bf H}_{bs}{\bf{\Phi }}_i{\bf{h}}_{sa_1}\right)a_{1,i_1}+{\bf{z}}_{b}^{(i_1)},\label{y_b}\vspace{-0mm}
\end{equation}
\begin{equation}
{{{y}}_{2}^{(i_1)}} =\sqrt{p_1}\left( {{h}}_{a_1 a_2}+{\bf{h}}_{sa_2}^T{\bf{\Phi }}_i{\bf{h}}_{sa_1} \right)a_{1,i_1}  + {z_2^{(i_1)}},\label{y_2}\vspace{-0mm}
\end{equation}
where $p_1$ denotes the transmit power of each pilot symbol, ${\mathbf{\Phi }}_{i_1} = \text{diag}\left({\bf{v}}_{i_1} \right)$ denotes the corresponding reflection coefficient matrix, ${\bf{z}}_{b}^{(i_1)} \!\sim\! {\mathcal{CN}} \left( {{\bf 0},\sigma^2{\bf I}} \right)$ denotes the independent CSCG noise vector at the BS and ${z_2^{(i_1)}} \!\sim\! {\mathcal{CN}} \left( {0,\sigma^2} \right)$ denotes the CSCG noise at A2. Let  ${\bf{\tilde{v}}}_{i_1}\!=\![1, {\bf{v}}_{i_1}^T]^T$, ${\bf{\tilde{H}}}_{ba_1}=[{\bf{h}}_{ba_1}, {\bf{H}}_{bsa_1}]$, and ${\bf{\tilde{h}}}_{a_1a_2}\!=\![h_{a_1a_2}, {\bf{h}}_{a_1sa_2}]$. By simply setting the pilot symbols as $a_{1,i_1}\!=\!1$, $\forall i_1$, (\ref{y_b}) and (\ref{y_2}) are rewritten as
\begin{equation}
{\bf{y}}_b^{(i_1)}=\sqrt{p_1}{\bf{\tilde{H}}}_{ba_1} {{\bf{\tilde{v}}}_{i_1}} +{\bf{z}}_{b}^{(i_1)},\vspace{-0mm}
\end{equation}
\begin{equation}
{{{y}}_{2}^{(i_1)}} =\sqrt{p_1} {\bf{\tilde{h}}}_{a_1a_2} {{\bf{\tilde{v}}}_{i_1}}+ {z_2^{(i_1)}}.\vspace{-0mm}
\end{equation}
By stacking the received signal vectors $\{{\bf{y}}_b^{(i_1)}\}_{i_1=1}^{\tau_1}$ and signals  $\{{{y}}_2^{(i_1)}\}_{i_1=1}^{\tau_1}$ into ${{\bf{Y}}_{b,1}}=[{\bf{y}}_b^{(1)},...,{\bf{y}}_b^{(\tau_1)}]$ and ${{\bf{y}}_2}=[{{y}}_2^{(1)},...,{{y}}_2^{(\tau_1)}]$, respectively, we obtain
\begin{equation}
\label{Y}
{{\bf{Y}}_{b,1}}=\sqrt{p_1}{\bf{\tilde{H}}}_{ba_1} {{\bf{\tilde{V}}}_1} +{\bf{Z}}_{b,1},
\end{equation}
\begin{equation}
\label{y_a2}
{{\bf{y}}_{2}} =\sqrt{p_1}{\bf{\tilde{h}}}_{a_1a_2} {{\bf{\tilde{V}}}_1} + {{\bf{z}}_2},
\end{equation}  
where ${{\bf{\tilde{V}}}_1}$, ${\bf{Z}}_{b,1}$ and ${{\bf{z}}_2}$ are defined as ${{\bf{\tilde{V}}}_1}=[{{\bf{\tilde{v}}}_1},...,{{\bf{\tilde{v}}}_{\tau_1}}]$,  ${\bf{Z}}_{b,1}=[{\bf{z}}_{b}^{(1)},...,{\bf{z}}_{b}^{(\tau_1)}]$ and ${{\bf{z}}_2}=[{z_2^{(1)}},...,{z_2^{(\tau_1)}}]$, respectively. By properly constructing ${{\bf{\tilde{V}}}_1}$ such that $\text{rank}({{\bf{\tilde{V}}}_1})=N+1$, the least-square (LS) estimations of ${\bf{\tilde{H}}}_{ba_1}$ and ${\bf{\tilde{h}}}_{a_1a_2}$ based on (\ref{Y}) and (\ref{y_a2}) are respectively given by\vspace{-0.5mm}
\begin{equation}
\label{H_ba1}
{\bf{\hat{H}}}_{ba_1}=[{\bf{\hat h}}_{ba_1}, {\bf{\hat H}}_{bsa_1}]=\frac{1}{\sqrt{p_1}}{{\bf{Y}}_{b,1}} {{\bf{\tilde{V}}}_1^{H}}{({\bf{\tilde{V}}}_1{\bf{\tilde{V}}}_1^{H})^{-1}},\vspace{-0mm}
\end{equation}
\begin{equation}
\label{h_a1a2}
{\bf{\hat{h}}}_{a_1a_2} =[{\hat h}_{a_1a_2}, {\bf{{\hat h}}}_{a_1sa_2}]= \frac{1}{\sqrt{p_1}}{{\bf{y}}_{2}} {{\bf{\tilde{V}}}_1^{H}}{({\bf{\tilde{V}}}_1{\bf{\tilde{V}}}_1^{H})^{-1}}.
\vspace{-0mm}
\end{equation}
After that, A2 feeds the estimated ${\bf \hat h}_{a_1sa_2}$ back to the BS.\footnote{For simplicity, we assume all channel feedbacks in this paper are lossless and error-free.}

Note that $\tau_1\ge N+1$ is required to satisfy the condition $\text{rank}({{\bf{\tilde{V}}}_1})=N+1$, and a general design of ${{\bf{\tilde{V}}}_1}$ is given in \cite{Jensen}. By setting $\tau_1=N+1$ to minimize the training overhead, ${{\bf{\tilde{V}}}_1}$ can be constructed based on the discrete Fourier transform (DFT) matrix with dimension $N+1$ \cite{zheng}, i.e., \vspace{-0mm}
\begin{equation}\small
\label{V_tilde}
{{\bf{\tilde{V}}}_1}=\left[
\begin{aligned}
&1~~~&~1~~~~~~~                   &...~~&~1~~~~~                   \\
&1~~~&e^{-j\frac{2\pi}{N+1}}~~~   &...~~&e^{-j\frac{2\pi }{N+1}N}  \\
&.   &~.~~~~~~~                   &...~~&~.~~~~~                   \\
&1~~~&e^{-j\frac{2\pi}{N+1}N}~~~  &...~~&e^{-j\frac{2\pi }{N+1}N^2} 
\end{aligned}
\right].\vspace{1mm}
\end{equation}
In this case,  ${{\bf{\tilde{V}}}_1^{H}}{({\bf{\tilde{V}}}_1{\bf{\tilde{V}}}_1^{H})^{-1}}$ can be efficiently computed by  ${{\bf{\tilde{V}}}_1^{H}}{({\bf{\tilde{V}}}_1{\bf{\tilde{V}}}_1^{H})^{-1}}=\frac{1}{N+1}{{\bf{\tilde{V}}}_1^H}$.

Next, A2 transmits pilot symbol $a_{2,i_2}$ with power $p_2$, $i_2=\tau_1+1,...,\tau_1+\tau_2$. Thus, the received signals at the BS are given by{\footnote{Here, the channel feedback from A1 to the BS is omitted for simplicity although A1 can help improve the  estimate of ${\bf h}_{a_1sa_2}$ based on the pilots from A2.}}
\begin{equation}
{\bf{y}}_b^{(i_2)}=\sqrt{p_2}{\bf{\tilde{H}}}_{ba_2} {{\bf{\tilde{v}}}_{i_2}} {a_{2,i_2}} +{\bf{z}}_{b}^{(i_2)},\vspace{-0mm}
\end{equation}
where ${\bf{\tilde{H}}}_{ba_2}=[{\bf{h}}_{ba_2}, {\bf{H}}_{bsa_2}]$. By setting $a_{2,i_2}=1$, $\forall i_2$, and stacking the received signal vectors $\{{\bf{y}}_b^{(i_2)}\}_{i_2=\tau_1+1}^{\tau_1+\tau_2}$ into ${{\bf{Y}}_{b,2}}=[{\bf{y}}_b^{(\tau_1)},...,{\bf{y}}_b^{(\tau_1+\tau_2)}]$, we obtain
\begin{equation}
\label{Y_2}
{{\bf{Y}}_{b,2}}=\sqrt{p_2}{\bf{\tilde{H}}}_{ba_2} {{\bf{\tilde{V}}}_2} +{\bf{Z}}_{b,2},
\end{equation}
where ${{\bf{\tilde{V}}}_2}$ and ${\bf{Z}}_{b,2}$ are defined as ${{\bf{\tilde{V}}}_2}=[{{\bf{\tilde{v}}}_{\tau_1+1}},...,{{\bf{\tilde{v}}}_{\tau_1+\tau_2}}]$ and ${\bf{Z}}_{b,2}=[{\bf{z}}_{b}^{(\tau_1+1)},...,{\bf{z}}_{b}^{(\tau_1+\tau_2)}]$, respectively. Accordingly, the LS estimate of  ${\bf{\hat{H}}}_{ba_2}$ is given by \vspace{-0mm}
\begin{equation}
\label{H_ba2}
{\bf{\hat{H}}}_{ba_2}=[{\bf{\hat h}}_{ba_2}, {\bf{\hat H}}_{bsa_2}]=\frac{1}{\sqrt{p_2}}{{\bf{Y}}_{b,2}} {{\bf{\tilde{V}}}_2^{H}}{({\bf{\tilde{V}}}_2{\bf{\tilde{V}}}_2^{H})^{-1}}.\vspace{-0mm}
\end{equation}
Similarly, $\tau_2 \ge N+1$ is required to satisfy $\text{rank}({{\bf{\tilde{V}}}_2})=N+1$. By setting $\tau_2=N+1$,  ${{\bf{\tilde{V}}}_2}$ can also be constructed as (\ref{V_tilde}).

Now, with ${\bf \hat h}_{a_1sa_2}$ fed back from A2, the BS acquires the estimated BS-IRS-A1, BS-IRS-A2 and A1-IRS-A2 channels, which are given by
\begin{equation}
{\bf{\hat H}}_{bsa_1}={\bf{\hat H}}_{bs}{\rm{diag}}({\bf{\hat h}}_{sa_1}),~ {\bf{\hat H}}_{bsa_2}={\bf{\hat H}}_{bs}{\rm{diag}}({\bf{\hat h}}_{sa_2}),~ \text{and } {\bf{\hat h}}_{a_1sa_2}={\bf{\hat h}}_{sa_2}^T{\rm{diag}}({\bf{\hat h}}_{sa_1}).\label{H_bra1}
\end{equation}
Based on (\ref{H_bra1}), the BS computes ${\bf{\hat H}}_{bs}\odot{\bf{\hat H}}_{bs}$ as ${\bf{\hat H}}_{bs}\odot{\bf{\hat H}}_{bs}={\bf{\hat H}}_{bsa_1} \odot {\bf{\hat H}}_{bsa_2}({\rm{diag}}({\bf{\hat h}}_{a_1sa_2}))^{-1}$. By defining ${\bf{G}}={\bf{\hat H}}_{bsa_1} \odot {\bf{\hat H}}_{bsa_2}({\rm{diag}}({\bf{\hat h}}_{a_1sa_2})^{-1})$, ${\bf{\hat H}}_{bs}\odot{\bf{\hat H}}_{bs}$ is rewritten as
\begin{equation}\small
\label{H_br}
{\bf{\hat H}}_{bs}\odot{\bf{\hat H}}_{bs}
=\left[
\begin{aligned}
&{\bf{G}}(1,1)&{\bf{G}}(1,2)~~~&...&{\bf{G}}(1,N)~\\
&{\bf{G}}(2,1)&{\bf{G}}(2,2)~~~&...&{\bf{G}}(2,N)~\\
&~~.          &.~~~~~          &...&.~~~                \\
&{\bf{G}}(M,1)&{\bf{G}}(M,2)~~ &...&{\bf{G}}(M,N)
\end{aligned}
\right],
\end{equation}
where ${\bf G}(m,n)\!=\!{\bf \hat H}_{bs}^2(m,n)$. By letting $g_{mn}=\sqrt{|{\bf{G}}(m,n)|}e^{j \frac{\angle {\bf{G}}(m,n)}{2}}$, each element in ${\bf{\hat H}}_{bs}$ is obtained as\vspace{-1.5mm} 
\begin{equation}
\label{H_bs}
 {\bf{\hat H}}_{bs}(m,n)=\pm g_{mn}, \forall m, n, \vspace{-1mm}
\end{equation}
i.e., we have recovered each ${\bf{\hat H}}_{bs}(m,n)$ but with  a +/- sign uncertainty. However, it will be unveiled later in Section III-B that such partial CSI is sufficient for resolving the cascaded BS-IRS-user channels ${\bf H}_{bsu_k}$'s of all users without ambiguity. 

To summarize, channel estimation Phase I consists of the following four steps:
\begin{itemize}
	\item A1 transmits pilot symbols while the BS and A2 estimate ${\bf{H}}_{bsa_1}$ and ${\bf{h}}_{a_1sa_2}$, respectively.
	\item A2 feeds ${\bf \hat {h}}_{a_1sa_2}$ back to the BS.
	\item A2 transmits pilot symbols while the BS estimates ${\bf{H}}_{bsa_2}$.
	\item The BS computes ${\bf{\hat  H}}_{bs}\odot{\bf{\hat H}}_{bs}$ based on ${\bf{\hat H}}_{bsa_1}$, ${\bf{\hat h}}_{a_1sa_2}$ and ${\bf{\hat H}}_{bsa_2}$.
\end{itemize}

\vspace{-2.5mm}
\subsection{Phase II: Estimation of ${\bf{h}}_{bu_k}$ and  ${\bf{H}}_{bsu_k}$}\vspace{-0.5mm}
As shown in Fig. 2, in each channel estimation Phase II, the BS first estimates the BS-user channels. Then, by leveraging the estimation of ${\bf{H}}_{bs}\odot{\bf{H}}_{bs}$ obtained in channel estimation Phase I, the IRS-user channels are estimated. Combining the estimated CSI of the BS-IRS link and IRS-user links, the cascaded BS-IRS-user channels are finally obtained.

Specifically, to estimate the direct BS-user channels, the IRS is first turned off. Denoting the pilot symbol transmitted by user $k$ by $x_{k,i_3}$, where $i_3=1,...,\tau_3$, then $\tau_3 \ge K$ pilot symbols are sufficient to estimate ${\bf h}_{bu_k}$, $\forall k$ \cite{Liu}, of which the details are omitted for brevity.

Next, all of the IRS elements are turned on to reflect signals and the IRS-user channels are estimated with properly designed reflection coefficients and pilot symbols. By denoting the pilot symbol transmitted from user $k$ as $x_{k,i_4}$, with transmit power $p$, where $i_4=\tau_3+1,...,\tau_3+\tau_4$, the received signals at the BS are given by\vspace{-1.5mm}
\begin{equation}
{\bf{y}}_b^{(i_4)}=\sqrt{p}\sum_{k=1}^{K}\left({\bf{h}}_{bu_k}+{\bf H}_{bs}{\bf{\Phi }}_{i_4}{\bf{h}}_{su_k}\right)x_{k,i_4}+{\bf{z}}_{b}^{(i_4)}.\label{y_b_1}\vspace{-1mm}
\end{equation}
Let ${\bf H}_{bs}{\bf{\Phi }}_{i_4}{\bf{h}}_{su_k}\!=\!{\bf{ H}}_{bsu_k}{\bf{v }}_{i_4}$, where ${\bf H}_{bsu_k}\!=\!{\bf H}_{bs}\text{diag}({\bf h}_{su_k})$ and ${\bf{\Phi }}_{i_4}=\text{diag}({\bf{v}}_{i_4})$. Then, by removing the signals from the direct channels based on the estimated ${\bf{\hat h}}_{bu_k}$'s, ${\bf{y}}_b^{(i_4)}$ can be re-expressed as\vspace{-1.5mm}
\begin{equation}
{\bf{\bar y}}_b^{(i_4)}=\sqrt{p}\sum_{k=1}^{K}{\bf{ H}}_{bsu_k}{\bf{v }}_{i_4}x_{k,{i_4}}+{\bf{\bar z}}_{b}^{({i_4})},\label{y_b_2}\vspace{-1mm}
\end{equation}
where  $
{\bf{\bar z}}_{b}^{({i_4})}\!=\!{\bf{z}}_{b}^{({i_4})}+\sqrt{p}\sum_{k=1}^{K}\left({\bf{h}}_{bu_k}\!-\!{\bf{\hat h}}_{bu_k}\right)x_{k,{i_4}}\notag\vspace{-1mm}$ 
is the effective noise due to both the receiver noise and the channel estimation error in ${\bf{\hat h}}_{bu_{k}}$. Different from the conventional schemes \cite{yangyf, you, zheng,Jensen} where $N$ pilot symbols are required to estimate each ${\bf{H}}_{bsu_k}$, we propose to decouple the estimation of   ${\bf{H}}_{bsu_k}$ into estimating ${\bf{ H}}_{bs}$ (common to all users) and ${\bf{h}}_{su_k}$ by leveraging the partial CSI of ${\bf{ H}}_{bs}$ obtained in channel estimation Phase I, thus greatly reducing the real-time training overhead since the dimension of ${\bf{h}}_{su_k}$ is much lower than that of its corresponding ${\bf{H}}_{bsu_k}$. Next, we consider the following two cases, depending on whether $M \ge N$ or not.

{\bf{Case 1:}} $M\ge N$. 
In this case, users send pilot symbols consecutively to facilitate the BS to estimate ${\bf{h}}_{su_{k}}$  independently. Specifically, at each time slot $i_4$, only one user transmits pilot symbol $x_{k,i_4}$. Consider user $k$ and let ${\bf{\Phi }}_{i_4}={\bf{I}}$,  $x_{k,i_4}=1$ and $x_{k_1,i_4}=0$, $k_1 \ne k$. ${\bf{\bar y}}_b^{(i_4)}$ in (\ref{y_b_2}) is thus rewritten as
\begin{equation}
\label{y_est}
{\bf{\bar y}}_b^{(i_4)}=\sqrt{p}{\bf{ H}}_{bs}{\bf{h}}_{su_k}+{\bf{\bar z}}_{b}^{(i_4)},
\end{equation}
where ${\bf{\bar z}}_{b}^{(i_4)}\!=\!{\bf{z}}_{b}^{(i_4)}\!+\!\sqrt{p}\left({\bf{h}}_{bu_k}\!-\!{\bf{\hat h}}_{bu_k}\right)$ . 

Generally, given ${\bf{H}}_{bs}$, there are in total $N$ unknowns to be estimated in ${\bf{h}}_{su_k}$, while the BS has  $M$ observations from its $M$ receiving antennas over each pilot symbol. Thus, if $M \ge N$, only one pilot symbol is sufficient for estimating each IRS-user channel. However, there still exists a crucial problem that each entry in the estimated ${\bf{\hat H}}_{bs}$ has two possible values as shown in (\ref{H_bs}). Due to such +/- sign uncertainty, ${\bf{h}}_{su_k}$ cannot be uniquely estimated. To overcome this issue, we provide {\bf Lemma 1} and {\bf Proposition 1} in the next to show that the estimate of ${\bf H}_{bsu_k}$ will not be affected by such sign uncertainty and it can be uniquely obtained.  \vspace{-0mm}
\begin{lemma}
	Given ${\bf{ H}}_{bsa_1}$, ${\bf{H}}_{bs}(m,n)$ can be expressed as ${\bf{H}}_{bs}(m,n)\!=\!\alpha_{mn}{\bf{ H}}_{bs}(1,n)$, where $\alpha_{mn}\!=\!\frac{{\bf{ H}}_{bsa_1}(m,n)}{{\bf{ H}}_{bsa_1}(1,n)}$, $\forall m,n$. 
	
	\begin{myproof}
		Since ${\bf{ H}}_{bsa_1}(1,n)={{\bf{ H}}_{bs}(1,n)}{{\bf{ h}}_{sa_1}(n)}$ and ${\bf{ H}}_{bsa_1}(m,n)={{\bf{ H}}_{bs}(m,n)}{{\bf{ h}}_{sa_1}(n)}$, we have ${\bf{ H}}_{bs}(m,n)=\frac{{\bf{ H}}_{bsa_1}(m,n)}{{\bf{ H}}_{bsa_1}(1,n)}{\bf{ H}}_{bs}(1,n)$, which thus completes the proof.  
	\end{myproof}
\end{lemma}
\vspace{-2mm}
Lemma 1 reveals that for given ${\bf{H}}_{bsa_1}$, there are $N$ rather than $MN$ unknowns in  ${\bf{H}}_{bs}$ and it can be rewritten as\vspace{-3mm}
\begin{equation*}\small
\label{H_bs_2}
{\bf{ H}}_{bs}\!=\!\left[
\begin{aligned}
&~~{{\bf  { H}}_{bs}(1,1)}         &{{\bf{ H}}_{bs}(1,2)}~~         &~~...&{{\bf{ H}}_{bs}(1,N)}~~\\
&\alpha_{21}{{\bf{H}}_{bs}(1,1)}&\alpha_{22}{{\bf{H}}_{bs}(1,2)}&~~...&\alpha_{2N}{{\bf{H}}_{bs}(1,N)}\\
&~~~~~~~.            &.~~~~~~~~           &~~...&.~~~~~~~~               \\
&\alpha_{M1}{{\bf{H}}_{bs}(1,1)}&\alpha_{M2}{{\bf{H}}_{bs}(1,2)}&~~...&\alpha_{MN}{{\bf{H}}_{bs}(1,N)}
\end{aligned}
\right].
\end{equation*}
In other words, once ${{\bf{H}}_{bs}(1,n)}$, $n=1,...,N$, and ${\bf{H}}_{bsa_1}$ are given, the whole matrix ${{\bf{H}}_{bs}}$ can be recovered uniquely.

With given  ${\bf{H}}_{bs}\odot{\bf{H}}_{bs}$ and ${\bf{H}}_{bsa_1}$, we assume that ${\bf G}(m,n)\!=\!{\bf H}_{bs}^2(m,n)$ and $g_{mn}\!=\!\sqrt{\!|{\bf{G}}(m,n)\!|}$ $e^{j \frac{\angle {\bf{G}}(m,n)}{2}}$,{\footnote{Here, we neglect the noise such that $\!{\bf{\hat H}}_{bs}\odot{\bf{\hat H}}_{bs}\!=\!{\bf{H}}_{bs}\odot{\bf{H}}_{bs}\!$ for the convenience of showing the unique channel recovery under sign ambiguities.}} then ${{\bf{H}}_{bs}(1,n)}$ has a +/- sign ambiguity as ${{\bf{H}}_{bs}(1,n)}=\pm g_{1n}, \forall n$, which cannot be uniquely determined. However, motivated by Lemma 1, we can construct a candidate of ${\bf H}_{bs}$ by assuming that ${\bf \tilde H}_{bs}(1,n)=g_{1n}$, $\forall n$, which is given by
\begin{equation*}\small
\label{H_bs_tilde}
{\bf \tilde H}_{bs}=\left[
\begin{aligned}
&~~g_{11}         &g_{12}~~         &~~...&g_{1N}~~\\
&\alpha_{21}g_{11}&\alpha_{22}g_{12}&~~...&\alpha_{2N}g_{1N}\\
&~~~~.            &.~~~~~           &~~...&.~~~~~                \\
&\alpha_{M1}g_{11}&\alpha_{M2}g_{12}&~~...&\alpha_{MN}g_{1N}
\end{aligned}
\right].
\end{equation*}\vspace{-3mm}
Then, based on ${\bf{\tilde H}}_{bs}$, the LS estimate of ${\bf{h}}_{su_k}$ is given by
\begin{equation}
    \label{h_suk}
	{\bf{\tilde  h}}_{su_k}=\frac{1}{\sqrt{p}}({\bf{\tilde H}}_{bs}^H{\bf{\tilde H}}_{bs})^{-1}{\bf{\tilde H}}_{bs}^H{\bf{\tilde y}}_b^{(i_4)},
\end{equation}
where ${\bf{\tilde y}}_b^{(i_4)}=\sqrt{p}{\bf{ H}}_{bs}{\bf{h}}_{su_k}$ is the signal received at the BS by neglecting the noise. However, since ${\bf{\tilde H}}_{bs}(1,n)={\bf{ H}}_{bs}(1,n)$ may not always be true, it remain unknown whether ${\bf{\tilde h}}_{su_k} = {\bf{h}}_{su_k}$ or not. Next, we present the following proposition which shows the uniqueness of the estimate of the cascaded BS-IRS-user channels at the BS (by ignoring the noise).
\vspace{-3mm}
\begin{Prop}
	Regardless of whether ${\bf{H}}_{bs}={\bf{\tilde H}}_{bs}$ or not and assuming no noise, it always holds that\vspace{-3mm}
	\begin{equation}
	\label{H_bsuk_est}
	{\bf{H}}_{bsu_k}={\bf{\tilde H}}_{bs}\text{diag}({\bf{\tilde h}}_{su_k}) .\vspace{-3mm}
	\end{equation}
	
	\begin{myproof}
	Please refer to Appendix. 
	\end{myproof}\vspace{-3mm}
\end{Prop}

Based on Proposition 1, we can estimate ${\bf{ H}}_{bsu_k}$'s by exploiting the partial CSI ${\bf{\hat H}}_{bs}\odot{\bf{\hat H}}_{bs}$ obtained in channel estimation Phase I. Since each user only transmits one pilot symbol for estimating ${\bf{ H}}_{bsu_k}$, it takes at minimum $\tau_4=K$ pilot symbols to estimate all cascaded BS-IRS-user channels.

{\bf{Case 2:}} $M < N$. In this case, if  users transmit pilot symbols one by one, it takes at least $\left\lceil {\frac{N}{M}} \right\rceil $ pilot symbols for the BS to estimate each ${\bf{h}}_{su_k}$ and thus the total training overhead is $K\left\lceil {\frac{N}{M}} \right\rceil $, which is the same as that in \cite{Dai}. To reduce such overhead, we propose a scheme based on user-grouping and orthogonal pilot symbols/reflection coefficients with a lower training overhead of $\left\lceil  {\frac{KN}{M}} \right\rceil$, which consists of three steps described as follows.

{{Step 1: }}Divide the $K$ users into $L=\left\lceil  {\frac{K}{M}} \right\rceil$ groups such that there are $M$ users in each of the first $L-1$ groups and $M_1$ ($M_1\le M$) users in the last group, i.e., $K=(L-1)M+M_1$.

{{Step 2: }}For each of the first $L-1$ groups, there are in total $MN$ unknowns to be estimated in $M$ users' IRS-user channels, while the BS has $M$ observations. Take the first group as an example. By denoting the pilot symbol transmitted by user $k$ at time slot $i_{4,1}$ by $x_{k,i_{4,1}}$, where $k=1,..,M$ and $i_{4,1}=\tau_3+1,...,\tau_3+\tau_{4,1}$, then the pilot symbols can be expressed in a vector form as ${\bf{x}}_{i_{4,1}}=[x_{1,i_{4,1}},...,x_{M,i_{4,1}}]^T$. As a result, the received signal vector at the BS over $\tau_{4,1}$ pilot symbols (by removing those from the direct channels) can be written as 
\begin{equation}
\label{y_case2}
\!\underbrace {\left[ {\begin{array}{*{20}{c}}
		{{{\bf{\bar y}}^{(\tau_3+1)}}}\\
		{{{\bf{\bar y}}^{(\tau_3+2)}}}\\
		\begin{array}{l}
		.\\		
		\end{array}\\
		{{{\bf{\bar y}}^{\left( \tau_3+\tau_{4,1} \right)}}}
		\end{array}} \right]}_{{\bf \bar y}_{4,1}} \!=\!\sqrt{p} \underbrace {\left[ {\begin{array}{*{20}{c}}
		{\bf{x}}^T_{{\tau_3+1}}\otimes\left({{{\bf{H}}_{bs}}{{\bf{\Phi}}_{\tau_3+1}}}\right)	\vspace{1mm}\\ 
		{\bf{x}}^T_{{\tau_3+2}}\otimes\left({{{\bf{H}}_{bs}}{{\bf{\Phi}} _{\tau_3+2}}}\right)\\
		.\\		
		{\bf{x}}^T_{{\tau_3+\tau_{4,1}}}\otimes\left({{{\bf{H}}_{bs}}{{\bf{\Phi}} _{\tau_3+\tau_{4,1}}}}\right)
		\end{array}} \right]}_{{\bf{B}}_1 \in {\mathbb{C}^{M\tau_{4,1} \times MN}}} \underbrace {\left[ {\begin{array}{*{20}{c}}
		{{{\bf{h}}_{s{u_1}}}}\\
		{{{\bf{h}}_{s{u_2}}}}\\
		\begin{array}{l}
		.\\			
		\end{array}\\
		{{{\bf{h}}_{s{u_M}}}}
		\end{array}} \right]}_{{\bf h}_1\in {\mathbb{C}^{MN \times 1 }}}\!+\!\underbrace {\left[ {\begin{array}{*{20}{c}}
		{\bf{\bar z}}_{b}^{({\tau_3+1})}\vspace{1mm}\\
		{\bf{\bar z}}_{b}^{({\tau_3+2})}\\
		\begin{array}{l}
		.\\		
		\end{array}\\
		{\bf{\bar z}}_{b}^{({\tau_3+\tau_{4,1}})}
		\end{array}} \right]}_{{\bf z}_{4,1}}\!,
\end{equation}
where ${\mathbf{\Phi }}_{i_{4,1}} = \text{diag}\left( {\bf v}_{i_{4,1}} \right)$ denotes the reflection matrix. As such, by properly designing the training reflection coefficients $\{{\bf v}_{i_{4,1}}\}_{i_{4,1}=\tau_3+1}^{\tau_3+\tau_{4,1}}$ of the IRS and the pilot symbol vectors $\{{\bf x}_{i_{4,1}}\}_{i_{4,1}=\tau_3+1}^{\tau_3+\tau_{4,1}}$ of the $M$ users such that $\text{rank}({\bf B}_1)=MN$, the LS estimate of ${\bf h}_1$ is given by
\begin{equation}
\label{h_ruk_est_2}
{\bf \hat h}_1= \frac{1}{\sqrt{p}}{\left( {{{\bf{B}}_1^H}{\bf{B}}_1} \right)^{ - 1}}{{\bf{B}}_1^H}{\bf \bar y}_{4,1}.
\end{equation}
Noting that $\tau_{4,1} \ge N$ is required to ensure the condition ${\text{rank}}({\bf B}_1)=MN$. Specifically, by setting $\tau_{4,1}=N$ and denoting ${\bf X}_2=[{\bf x}_{\tau_3+1},...,{\bf x}_{\tau_3+N}]$, we can design the pilot symbol matrix transmitted by the $M$ users over the $\tau_{4,1}=N$ pilot symbols as
\begin{equation}\small
{\bf X}_2\!=\!\!\left[\! {\begin{array}{*{20}{c}}
	1&1&1&{...}&1\\
	1&{{e^{ - j\theta }}}&{{e^{ - j2\theta }}}&{...}&{{e^{ - j\left( {N - 1} \right)\theta }}}\\
	.&.&{}&.&{}\\
	1&{{e^{ - j\left( {M - 1} \right)\theta }}}&{{e^{ - j2\left( {M - 1} \right)\theta }}}&{...}&{{e^{ - j{\left( {M - 1} \right){\left( {N - 1} \right)}}\theta }}}
	\end{array}} \!\right]\!,\notag
\end{equation}
where $\theta=\frac{2\pi}{N}$. On the other hand, by denoting ${\bf V}_1=[{\bf v}_{\tau_3+1},...,{\bf v}_{\tau_3+N}]$, the reflection coefficients of the $N$ reflecting elements during the $\tau_{4,1}\!=\!N$ pilot symbols are given by
\begin{equation}\small
{\bf V}_1=\left[ {\begin{array}{*{20}{c}}
	1&1&1&{...}&1\\
	1&{{e^{ - j\theta }}}&{{e^{ - j2\theta }}}&{...}&{{e^{ - j\left( {N - 1} \right)\theta }}}\\
	.&.&.&...&{.}\\
	1&{{e^{ - j\left( {N - 1} \right)\theta }}}&{{e^{ - j2\left( {N - 1} \right)\theta }}}&{...}&{{e^{ - j{{\left( {N - 1} \right)^2}}\theta }}}
	\end{array}} \right].\notag
\end{equation}
It should be noted that the exact ${\bf B}_1$ in (\ref{h_ruk_est_2}) is unknown due to the fact that we only have the estimation of ${\bf{H}}_{bs}\odot {\bf{H}}_{bs}$, instead of ${\bf{H}}_{bs}$. Nevertheless, similar to {\bf Case 1}, by setting ${\bf{H}}_{bs}={\bf \tilde{H}}_{bs} $ in ${\bf B}_1$, the cascaded channels of the users in the first group can be uniquely estimated as
\begin{equation}
[{\bf{\hat H}}_{bsu_1},...,{\bf{\hat H}}_{bsu_M}]={\bf{B}}_1{\rm{diag}}({\bf{\hat h}}_{1}).\vspace{-1mm}
\end{equation}
Similarly, the cascaded channels for the users in each of the following $L-2$ groups can be estimated, and the minimum training overhead is respectively given by $\tau_{4,l}=\tau_{4,1}=N$, where $l=2,...,L-1$.

{{Step 3: }}For the users in the last group, we assume that another $\tau_{4,L}$ pilot symbols are used to estimate the cascaded channels. Specifically, we denote the pilot symbol transmitted by user $k$ at time slot $i_{4,L}$ by $x_{k,i_{4,L}}$, where $k\!=\!(L\!-\!1)M\!+\!1,...,K$ and $i_{4,L}\!=\!\tau_3\!+\!(L\!-\!1)N\!+\!1,...,\tau_3\!+\!(L\!-\!1)N\!+\!\tau_{4,L}$. By setting $\bar K\!=\!(L\!-\!1)M$ and $\bar \tau\!=\!\tau_3+(L\!-\!1)N$, the pilot symbols can be expressed in a vector form as ${\bf{x}}_{i_{4,L}}=[x_{\bar K+1,i_{4,L}},...,x_{K,i_{4,L}}]^T$, while the reflection matrix can be written as ${\mathbf{\Phi }}_{i_{4,L}}\! =\! \text{diag}\left( {\bf v}_{i_{4,L}} \right)$, where $i_{4,L}\!=\!\bar \tau\!+\!1,...,\bar \tau\!+\!\tau_{4,L}$. Thus, the received signal vector at the BS over the $\tau_{4,L}$ pilot symbols (by removing those from the direct channels) is written as 
\begin{equation*}\small
\label{y_case2_2}
\underbrace {\left[ {\begin{array}{*{20}{c}}
		{{{\bf{\bar y}}^{(\bar \tau +1)}}}\\
		{{{\bf{\bar y}}^{(\bar \tau +2)}}}\\
		\begin{array}{l}
		.\\
		.
		\end{array}\\
		{{{\bf{\bar y}}^{\left( \bar \tau +\tau_{4,L} \right)}}}
		\end{array}} \right]}_{{\bf \bar y}_{4,L}} \!=\!\sqrt{p} 
\underbrace{ \left[ {\begin{array}{*{20}{c}}
		{\bf{x}}^T_{{\bar \tau+1}}\otimes\left({{{\bf{H}}_{bs}}{{\bf{\Phi}}_{\bar \tau+1}}}\right)	\\
		{\bf{x}}^T_{{\bar \tau+2}}\otimes\left({{{\bf{H}}_{bs}}{{\bf{\Phi}} _{\bar \tau+2}}}\right)\\
		.\\
		.\\
		{\bf{x}}^T_{{\bar \tau+\tau_{4,L}}}\otimes\left({{{\bf{H}}_{bs}}{{\bf{\Phi}} _{\bar \tau+\tau_{4,L}}}}\right)
		\end{array}} \right] }_{{\bf B}_2\in {\mathbb{C}}^{M\tau_{4,L}\times M_1N}}
\underbrace {\left[ {\begin{array}{*{20}{c}}
		{{{\bf{h}}_{s{u_{\bar K+1}}}}}\\
		{{{\bf{h}}_{s{u_{\bar K+2}}}}}\\
		\begin{array}{l}
		.\\
		.
		\end{array}\\
		{{{\bf{h}}_{s{u_{K}}}}}
		\end{array}} \right]}_{{\bf h}_L\in {\mathbb{C}}^{M_1N \times 1}}\!+\!
\underbrace{\left[ {\begin{array}{*{20}{c}}
		{\bf{\bar z}}_{b}^{({\bar \tau+1})}\\
		{\bf{\bar z}}_{b}^{({\bar \tau+2})}\\
		\begin{array}{l}
		.\\
		.
		\end{array}\\
		{\bf{\bar z}}_{b}^{({\bar \tau+\tau_{4,L}})}
		\end{array}} \right]}_{{\bf z}_{4,L}}.\vspace{-0mm}
\end{equation*}
As such, by properly designing the training reflection matrix of the IRS and the pilot symbols of users such that $\text{rank}({\bf B}_2)=M_1N$, the LS estimate of  ${\bf h}_L$ is given by\vspace{-0.5mm}
\begin{equation}
\label{h_ruk_est_3}
{\bf \hat h}_L= \frac{1}{\sqrt{p}}{\left( {{{\bf{B}}_2^H}{\bf{B}}_2} \right)^{ - 1}}{{\bf{B}}_2^H}{\bf \bar y}_{4,L}.\vspace{-0.5mm}
\end{equation}
Since $M\tau_{4,L} \ge M_1N$ is required to ensure the condition of  $\text{rank}({\bf B}_2)=M_1N$ with $\tau_{4,L}$ being an integer, we have $\tau_{4,L} \ge \left\lceil  {\frac{M_1N}{M}} \right\rceil$. Specifically, by setting $\tau_{4,L} = \left\lceil {\frac{M_1N}{M}} \right\rceil=N_1$, ${\bf X}_3=[{\bf x}_{\bar \tau+1},...,{\bf x}_{\bar \tau+N_1}]$ and ${\bf V}_2=[{\bf v}_{\bar \tau+1},...,{\bf v}_{\bar \tau+N_1}]$, ${\bf X}_3$ and ${\bf V}_2$ are respectively given by
\begin{equation*}\small
{\bf X}_3\!=\!\!\left[\! {\begin{array}{*{20}{c}}
	1&1&{...}&1\\
	1&{{e^{ - j\theta }}}&{...}&{{e^{ - j\left( {N_1 - 1} \right)\theta }}}\\
	.&.&{...}&.\\
	1&{{e^{ - j\left( {M_1 \!-\! 1} \right)\theta }}}&{...}&{{e^{ - j{\left( {M_1 \!-\! 1} \right){\left( {N_1 \!-\! 1} \right)}}\theta }}}
	\end{array}} \!\!\right]\!\!~, 
{\bf V}_2\!=\!\left[ {\begin{array}{*{20}{c}}
	1&1&{...}&1\\
	1&{{e^{ - j\theta }}}&{...}&{{e^{ - j\left( {N_1 - 1} \right)\theta }}}\\
	.&.&{...}&.\\
	1&{{e^{ - j\left( {N\! -\! 1} \right)\theta }}}&{...}&{{e^{ - j{{\left( {N - 1} \right)\left( {N_1 \!-\! 1} \right)}}\theta }}}
	\end{array}} \right].
\end{equation*}
Similarly, by setting ${\bf{H}}_{bs}={\bf \tilde{H}}_{bs}$ in ${\bf B}_2$, the cascaded channels of the users in the last group can be estimated as\vspace{-1mm}
\begin{equation}
\label{H_bsu_L}
[{\bf{\hat H}}_{bsu_{\bar K+1}},...,{\bf{\hat H}}_{bsu_K}]={\bf{B}}_2{\rm{diag}}({\bf{\hat h}}_{L}).\vspace{-1mm}
\end{equation}
In summary, the minimum training overhead for estimating the cascaded BS-IRS-user channels for {\bf Case 2} is  $\tau_4=(L-1)\tau_{4,1}+\tau_{4,L}=\left\lceil  {\frac{M(L-1)N+M_1N}{M}} \right\rceil=\left\lceil  {\frac{KN}{M}} \right\rceil$. 

Based on the above, the proposed Scheme 1 for estimating the BS-IRS-user cascaded channels ${\bf{\hat H}}_{bsu_k}$'s is summarized as Algorithm 1.

\setlength{\intextsep}{0pt}
\setlength{\textfloatsep}{2mm} 
\begin{algorithm}[t]
	\DontPrintSemicolon
	\LinesNumbered	
	\SetAlgoSkip{smallskip}
	\SetAlgoHangIndent{0.1em}
	\caption{Estimating ${\bf{H}}_{bsu_k}$ by Proposed Scheme 1}
	\label{Algorithm1}	
	\KwIn{${\bf{\hat H}}_{bs}\odot{\bf{\hat H}}_{bs}$, ${\bf{\hat H}}_{bsa_1}$.}
	\KwOut{${\bf{\hat H}}_{bsu_k}$, $\forall k$.}
	Define an $M \!\times\! N$ matrix $\bf G$ where ${\bf G}(m,n)\!=\!{\bf \hat H}_{bs}^2(m,n)$, $\forall m,n$. \\
	Let $g_{1n}=\sqrt{|{\bf{G}}(1,n)|}e^{j \frac{\angle {\bf{G}}(1,n)}{2}}$, $\forall n$.\\
	Define an $M \times N$ matrix ${\bf \tilde{H}}_{bs}$ where ${\bf \tilde{H}}_{bs}(1,n)=g_{1n}$ and ${\bf \tilde{H}}_{bs}(m,n)=\frac{{\bf{\hat H}}_{bsa_1}(m,n)}{{\bf{\hat H}}_{bsa_1}(1,n)}g_{1n}$, $\forall m,n$, and obtain ${\bf B}_1$ and ${\bf B}_2$ by setting ${\bf{H}}_{bs}={\bf \tilde{H}}_{bs}$.\\
	
		\eIf{$M \ge N$}
			{
			 \Repeat{{\rm $k=K$.}}
				{
				 User $k$ transmits $\!x_{k,i_4}\!$ and the BS obtains $\!{\bf{\bar y}}_b^{(i_4)}\!$ as in (\ref{y_est}).\\
			     BS estimates ${\bf{h}}_{su_k}$ as			    
		         $~~~~~{\bf{\hat h}}_{su_k}\!=\!\frac{1}{\sqrt{p}}({\bf \tilde{H}}_{bs}^H{\bf \tilde{H}}_{bs})^{-1}{\bf \tilde{H}}_{bs}^H{\bf{\bar y}}_b^{(i_4)}$.\\
		         BS estimates ${\bf{H}}_{bsu_k}$ as 
		    	 ${\bf{\hat H}}_{bsu_k}={\bf \tilde{H}}_{bs}{\rm{diag}}({\bf{\hat h}}_{su_k})$.\\
		    	 Update $k=k+1$. 
	        	}	
            }
    	    {
    	     Divide the $K$ users into $L=\left\lceil  {\frac{K}{M}} \right\rceil$ groups, set $l=1$.\\
    	     \Repeat{{\rm $l=L-1$.}}
	    	    { 
	    	     The users in the $l$-th group transmit pilot symobls ${\bf X}_2$ and the BS receives ${\bf \bar y}_{4,l}$.\\
	    	     BS estimates ${\bf \hat h}_1$ as ${\bf \hat h}_l= \frac{1}{\sqrt{p}}{\left( {{{\bf{B}}_1^H}{\bf{B}}_1} \right)^{ - 1}}{{\bf{B}}_1^H}{\bf \bar y}_{4,l}$.\\
	    	     BS estimates the cascaded channels as $[{\bf{\hat H}}_{bsu_{(l-1)M+1}},...,{\bf{\hat H}}_{bsu_{lM}}]={\bf{B}}_1{\rm{diag}}({\bf{\hat h}}_{l})$.	\\
	    	     Update $l=l+1$.	    	    	
	    	    }	
    	        The users in the last group transmit pilot symbols ${\bf X}_3$ and the BS receives ${\bf \bar y}_{4,L}$.\\
    	        BS estimates ${\bf \hat h}_L$ as (\ref{h_ruk_est_3}).\\
    	        BS estimates the cascaded channels as (\ref{H_bsu_L}).	
	        }	

	\vspace{-1mm}
\end{algorithm}

\vspace{-2mm}
\subsection{Training and Feedback Complexity}\vspace{-0mm}
\subsubsection{Training Overhead}
As discussed in Section III-A, the minimum training overhead for channel estimation Phase I is $\tau_1+\tau_2=2(N+1)$. Note that Phase I is executed only once during each channel coherence time $T_c$. In contrast, Phase II for estimating the cascaded BS-IRS-user channels needs to be conducted periodically during the remaining time of $T_c$, i.e., over each channel coherence time $T_u$, which results in a minimum training overhead equal to $\tau_3+\tau_4=2K$ for the case $M \!\ge\! N$ and $\tau_3+\tau_4=K+\left\lceil\frac{KN}{M}\right\rceil\!$ for the case $M \!<\! N$.  
\subsubsection{Feedback Complexity} 
In the proposed Scheme 1, the anchor node A2 needs to feed back ${\bf \hat h}_{a_1sa_2}$ to the BS, which is of size $N$, for computing ${\bf \hat {H}}_{bs}\odot{\bf \hat {H}}_{bs}$. This feedback of ${\cal O} (N)$ occurs only once during each channel coherence time $T_c$.

\vspace{-0mm}
\begin{Rem}
	Although our proposed Scheme 1 exploits the same fact that all cascaded BS-IRS-user channels share the common BS-IRS channel as those in \cite{Liu} and \cite{Dai}, they have significant differences as follows. For the scheme in \cite{Liu}, a reference user's cascaded channel is estimated first while the other users' cascaded channels are scaled versions of it, which thus can be recovered by estimating only the scaling parameters with much lower dimensions than their cascaded channels. However, since the cascaded channel estimation of all the other users relies on that of the reference user, there are two drawbacks in practice: (1) the reference user's channel needs to be updated once it changes, which thus increases the training overhead; and (2) the estimation accuracy of any other user's cascaded channel is sensitive to that of the reference user, which however is constrained by the limited power at the user side. In contrast, in the proposed Scheme 1, all cascaded channels are obtained from the BS-IRS channel estimated in channel estimation Phase I and IRS-user channels estimated in channel estimation Phase II. As compared to the scheme in \cite{Liu}, Phase I for estimating the BS-IRS channel only needs to be conducted once for each $T_c$, which thus reduces the training overhead. Moreover, the estimation can also be made sufficiently accurate since the anchors (e.g., IRS controllers) generally have more transmit power than  users to send the pilot signals and can be deployed properly for achieving strong BS-A1/A2 and IRS-A1/A2 links. Furthermore, given sufficiently accurate estimation of the BS-IRS channel, the  estimation of IRS-user channels in Phase II can achieve higher accuracy as well due to the short IRS-user distances. Note that although the scheme in \cite{Dai} also exploits the common BS-IRS channel to reduce training overhead, it incurs higher complexity and achieves much lower accuracy than Scheme 1. In particular, each BS antenna in \cite{Dai} is required to consecutively transmit pilot symbols during which the rest BS antennas estimate the corresponding BS-IRS-BS channel, thus incurring a training overhead of $M(N+1)$ in order to estimate the BS-IRS channel. However, due to the typically non-negligible self-interference in FD transceivers and significant double path loss (to/from IRS), estimating the BS-IRS-BS channel is much less accurate than estimating the BS-IRS-anchor channels (especially when the distance between the BS and IRS is much longer than that between the IRS and nearby anchors in practice), which thus renders the estimation of the desired BS-IRS-user channels in \cite{Dai} less accurate as well. Moreover, it is worth pointing out that the important +/- sign ambiguity issue raised in Section III-B is neglected in \cite{Dai}, which is solved rigorously in this paper. To summarize, our proposed Scheme 1 can achieve higher accuracy than the schemes proposed in \cite{Liu} and \cite{Dai}, but with reduced training overhead in general (see Table I later for a more detailed comparison).
\end{Rem}

\vspace{-0mm}
\section{Exploiting the Strong A1-IRS-A2 and A2-IRS-User Links}\vspace{-0mm}
As shown in Section III-C, the training overhead required in channel estimation Phase II of our proposed Scheme 1 crucially depends on the number of antennas at the BS, i.e., $M$ (similar to those proposed in \cite{Liu} and \cite{Dai}). Thus, it is practically desirable to devise an alternative anchor-assisted channel estimation scheme whose training overhead is not sensitive to $M$, such that the channel estimation efficiency can still be improved even with small or moderate values of $M$ (e.g., in WiFi system). In fact, given the estimation of the cascaded BS-IRS-A1 and A1-IRS-A2 channels obtained in Phase I of Scheme 1, the A2-IRS-user channels can be estimated with additional training by A2 and further exploited to facilitate the estimation of the cascaded BS-IRS-user channels at the BS.  Motivated by this, we propose in this section another anchor-assisted channel estimation scheme (termed Scheme 2) by exploiting the short distance/strong A1-IRS-A2 and A2-IRS-user channels.

\begin{figure*}[t]
	\centering
	\includegraphics[width=6.5in]{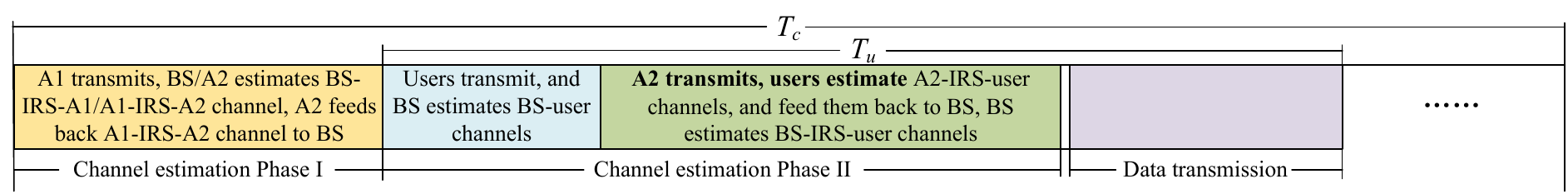}
	\caption{Channel estimation and data transmission protocol of the proposed anchor-assisted channel estimation Scheme 2.}\vspace{1mm}
\end{figure*}

For each channel coherence time $T_c$, the channel estimation and transmission protocol of our proposed Scheme 2 is shown in Fig. 3. First, the cascaded BS-IRS-A1 and A1-IRS-A2 channels are estimated separately in Phase I based on the pilot signals sent by A1. Note that different from Scheme 1, A2 is not required to transmit pilot signals in this phase. Then, for Phase II in each channel coherence interval $T_u$, after estimating the BS-user direct channels ${\bf h}_{bu_k}$, $\forall k$, A2 (instead of users in Scheme 1) transmits pilot symbols and each user estimates the cascaded A2-IRS-user channels independently. Based on the CSI of BS-IRS-A1 and A1-IRS-A2 links, which are obtained in Phase I as well as the feedback CSI of A2-IRS-user links estimated by the users in Phase II, the BS resolves the cascaded BS-IRS-user channels in real time. The details of the two channel estimation phases of Scheme 2 are described as follows.

\vspace{-2mm}
\subsection{Phase I: Estimation of  ${\bf{H}}_{bsa_1}$ and ${\bf{h}}_{a_1sa_2}$}
First, A1 transmits pilot symbols while the BS and A2 estimate ${\bf H}_{bsa_1}$ and ${\bf h}_{a_1sa_2}$, respectively. This part is the same as that in Scheme 1 and thus the details are omitted for brevity. Note that A2 is not required to transmit pilot symbols in this phase, thus the minimum training overhead is $N+1$, which is only half of that in Scheme 1.

\vspace{-2.5mm}
\subsection{Phase II: Estimation of ${\bf{h}}_{bu_k}$, ${\bf{h}}_{a_2su_k}$ and ${\bf{H}}_{bsu_k}$}\vspace{-0mm}

In this phase, the IRS is first turned off and the BS estimates the BS-use direct channels ${\bf{h}}_{bu_k}$'s, similar to Scheme 1. The minimum training overhead is thus $K$.

Next, the IRS is turned on to reflect signals. When A2 transmits pilot symbols, the users estimate their respective  A2-IRS-user channels simultaneously. Assuming that A2 transmits pilot symbol $a_{2,i}$ with power $p$, $i=1,...,{\tau}_0$, the received signal at user $k$ is expressed as{\footnote{For fair comparison, the transmit power of A2 in channel estimation Phase II is assumed to be identical as that of each user in Scheme 1, while in practice, A2 can transmit higher power than users for better estimation accuracy.}}\vspace{-0.2mm}
\begin{equation}
{{{y}}_{u_k}^{(i)}} =\sqrt{p} {\bf{\tilde{h}}}_{a_2u_k} {{\bf{\tilde{v}}}_{i}} {a_{2,i}}  + {z_{u_k}^{(i)}},\vspace{-0.3mm}
\end{equation}
where ${\bf{\tilde{v}}}_{i}=[1, {\bf{v}}_{i}^T]^T$,  ${\bf{\tilde{h}}}_{a_2u_k}=[h_{a_2u_k}, {\bf{h}}_{a_2su_k}]$, and ${z_{u_k}^{(i)}} \!\sim\! {\mathcal{CN}} \left( {0,\sigma^2} \right)$ denotes the CSCG noise at user $k$. By simply setting the pilot symbols as $a_{2,i}=1$, $\forall i$, and stacking the received signals  $\{{{{y}}_{u_k}^{(i)}}\}_{i=1}^{\tau_0}$ into  ${{\bf{y}}_{u_k}}=[{{y}}_{u_k}^{(1)},...,{{y}}_{u_k}^{(\tau_0)}]$, we obtain\vspace{-0.2mm}
\begin{equation}
\label{y_u_k}
{{\bf{y}}_{u_k}}=\sqrt{p}{\bf{\tilde{h}}}_{a_2u_k} {{\bf{\tilde{V}}}} + {{\bf{z}}_{u_k}},\vspace{-0.5mm}
\end{equation}   
where ${{\bf{\tilde{V}}}}=[{{\bf{\tilde{v}}}_1},...,{{\bf{\tilde{v}}}_{\tau_0}}]$ and ${{\bf{z}}_{u_k}}=[{z_{u_k}^{(1)}},...,{z_{u_k}^{(\tau_0)}}]$. By properly constructing ${{\bf{\tilde{V}}}}$ such that $\text{rank}({{\bf{\tilde{V}}}})=N+1$, the LS estimate of ${\bf{\tilde{h}}}_{a_2u_k}$ based on (\ref{y_u_k}) is given by\vspace{-0.5mm}
\begin{equation} 
\label{h_a2uk}
{\bf{\hat{h}}}_{a_2u_k} =[{\hat h}_{a_2u_k}, {\bf{{\hat h}}}_{a_2su_k}]= \frac{1}{\sqrt{p}}{{\bf{y}}_{u_k}}{{\bf{\tilde{V}}}^{H}}{({\bf{\tilde{V}}}{\bf{\tilde{V}}}^{H})^{-1}}.
\vspace{-0.5mm}
\end{equation}
Note that $\tau_0\ge N+1$ is required to satisfy the condition $\text{rank}({{\bf{\tilde{V}}}})=N+1$. By setting $\tau_0=N+1$, ${{\bf{\tilde{V}}}}$ can be constructed similarly as ${{\bf{\tilde{V}}}}_1$ in  (\ref{V_tilde}).

Next, each user feeds the estimated ${\bf{\hat h}}_{a_2su_k}$ back to the BS. With ${\bf{\hat H}}_{bsa_1}$ and ${\bf{\hat h}}_{a_1sa_2}$ estimated in  Phase I and ${\bf \hat{h}}_{a_2su_k}$ estimated in Phase II, the BS estimates the cascaded BS-IRS-user channel ${\bf{\hat H}}_{bsu_k}$ as
\begin{equation}
{\bf{\hat H}}_{bsu_k}={\bf{\hat H}}_{bsa_1} {\rm{diag}}({\bf{\hat h}}_{a_2su_k})({\rm{diag}}({\bf{\hat h}}_{a_1sa_2}))^{-1}.\label{H_bsu_S2}
\end{equation}

 \vspace{-5mm}
\subsection{Training and Feedback Complexity}
\subsubsection{Training Overhead}
Same as in Scheme 1, Phase I in Scheme 2 is also conducted only once. The difference is that the minimum training overhead is $N+1$ instead of $2(N+1)$ in Scheme 1 since only A1 is exploited to transmit pilot symbols for training. Moreover, the minimum training overhead for each Phase II is $K+\tau_0=K+N+1$, regardless of $M$.
\subsubsection{Feedback Complexity}
In Scheme 2, the feedback required is as follows. First, for  Phase I, A2 needs to feed back ${\bf\hat h}_{a_1sa_2}$ of size $N$ to the BS. Second, for Phase II, users need to feed back their individually estimated ${\bf{\hat h}}_{a_2su_k}$ to the BS, which results in the total feedback complexity of ${\cal O}(KN)$.

 \vspace{-2mm}
\subsection{Comparison of Proposed and Benchmark Schemes}
\subsubsection{Training Overhead}
The comparisons of the (minimum) channel training overhead (in terms of number of pilot symbols) and feedback complexity order between the proposed and two benchmark schemes are given in Table I. Note that the  estimation of BS-IRS-A1/A2 and A1-IRS-A2 channels in the proposed schemes and that of BS-IRS channel in \cite{Dai} are both executed once, while as shown in Table I, the  training overhead required by the proposed schemes is significantly less than that in \cite{Dai} (i.e., $2(N+1)$ and $N+1$ vs. $M(N+1)$, respectively). This is because in our two proposed schemes, either two anchors or only one anchor is required to transmit $2(N+1)$ and $N+1$ pilot symbols, respectively, in channel estimation Phase I, while in \cite{Dai} each of the $M$ BS antennas is required to transmit $N+1$ pilot symbols (termed channel estimation Phase I in \cite{Dai}). Next, for each channel estimation Phase II, we observe that if $K>N+1$, our proposed Scheme 2 outperforms Scheme 1 in terms of the training overhead, regardless of whether $M\ge N$ or not. Otherwise, if $K \le N$, for $M\ge N$, Scheme 1 always outperforms Scheme 2, while for $M < N$, Scheme 1 still outperforms Scheme 2 when  $\left\lceil\frac{KN}{M}\right\rceil<N+1$, i.e., $K \le M$. Considering that we usually have $K \ll M$ and $K \ll N$ in practice, Scheme 1 is more efficient than Scheme 2 in terms of training overhead. In addition, besides requiring the use of an FD BS, the scheme in \cite{Dai} also has a larger overhead than our proposed Scheme 1 in channel estimation Phase II when $M<N$ and $\text{mod}(N,M) \ne 0$, as well as Scheme 2 when $N+1 < K\left\lceil\frac{N}{M}\right\rceil$. For example, for $K=60$, $M=70$ and $N=80$, the training overheads in Phase II for Scheme 1, Scheme 2 and the scheme in \cite{Dai} are 129, 141 and 180 pilot symbols, respectively. Note that in the scheme proposed in \cite{Liu}, at least $K$ pilots and $N$ pilots are required to estimate all users' direct channels and the reference user's cascaded channel, respectively, which is termed channel estimation Phase I in Table. I and needs to be executed every $T_u$ during each channel coherence time $T_c$. In contrast, channel estimation Phase I in the other three schemes are all conducted only once for each $T_c$.

\begin{table*}[t]
	\footnotesize
	\renewcommand{\arraystretch}{1.8}
	\caption{Training Overhead   Comparison of Considered Schemes.}
	\label{Complexity}
	\centering 
		\begin{tabular}{|m{3.5cm}<{\centering}|m{1.8cm}<{\centering}|m{2cm}<{\centering}|m{1.8cm}<{\centering}|m{2.1cm}<{\centering}|m{2cm}<{\centering}|}
			\hline 
			\multirow{3}{*}{ } 
			&\multicolumn{2}{|c|}{Phase I } &\multicolumn{3}{|c|}{{Phase II }}
			\\
			\cline{2-6}       
			&\multirow{2}{*}{Overhead}
			&\multirow{2}{*}{Executed times}
			&\multicolumn{2}{|c|}{{Overhead}}
			&\multirow{2}{*}{Executed times}\\
			\cline{4-5}        &      &                    & $M \ge N$         &$M<N$     & \\
			\hline Proposed anchor-assisted Scheme 1            &$2(N+1)$                  &      1        &$2K$  &$K+\left\lceil {KN}/{M} \right\rceil$ &$\frac{T_c-2N-2}{T_u}$\\
			\hline Proposed anchor-assisted Scheme 2           &$N+1$                  &  1   &  \multicolumn{2}{|c|}{{$K+N+1$}} &$\frac{T_c-N-1}{T_u}$ \\
			\hline {{\cite{Liu}}} & $K+N$ & $\frac{T_c}{T_u}$& $K-1$ & $\left\lceil {(K-1)N}/{M} \right\rceil$ & $\frac{T_c}{T_u}$\\
			\hline {\cite{Dai}}   &$M(N+1)$  & 1  &$2K$  & $K+K\left\lceil {N}/{M} \right\rceil$ &$\frac{T_c-MN-M}{T_u}$ \\
			\hline
		\end{tabular} 
	\vspace{2mm}
\end{table*}

\subsubsection{Feedback Complexity}
It should be noted that both proposed schemes need CSI feedback from A2 to the BS in channel estimation Phase I, which is of ${\cal O}(N)$. Moreover, since we let A2 transmit pilot symbols for the users to estimate channels and feed them back to the BS in channel estimation Phase II of Scheme 2, a feedback complexity of ${\cal O}(KN)$ is required. In contrast, for both the benchmark schemes in \cite{Liu} and \cite{Dai}, the BS estimates the desired channels based on the pilots from the users, and thus no feedback is required.

\subsubsection{Advantages of Proposed Scheme 2 over Scheme 1} 
Despite that the proposed Scheme 1 may be superior to Scheme 2 in terms of the training overhead in Phase II (especially when $M$ is large) and feedback complexity, Scheme 2 could offer the following advantages over Scheme 1:  
\begin{itemize}
	\item First, the training overhead in channel estimation Phase I for Scheme 2 is half of that for Scheme 1.
	\item Second, in Scheme 2, the training overhead for estimating the cascaded BS-IRS-user channels in channel estimation Phase II is fixed as $N+1$, regardless of the value of $M$ or $K$. 
	\item Third, the users only feed back their estimated channels to the BS without the need of actively transmitting pilots for estimating the cascaded BS-IRS-user channels since A2 transmits pilots instead. This thus simplifies the training design from the users' perspective.  
	\item Last but not least, Scheme 2 can achieve a significantly higher accuracy than Scheme 1 for estimating the cascaded BS-IRS-user channels given their respective minimum training   overheads. This is because the former case exploits the much stronger A1-IRS-A2 and A2-IRS-user links rather than the (common) BS-IRS link in the latter case due to their drastically shorter link distances in practice, as will be verified and further discussed based on simulation results in Section V. 
\end{itemize}

\subsubsection{Performance under Perfect  Channel Estimation in Phase I} Given sufficiently large transmit power and/or sufficiently long training time, the estimation of BS-IRS-A1/A2 and A1-IRS-A2 channels in the proposed schemes and that of BS-IRS channel in \cite{Dai} tends to be error-free. In this case, the estimation accuracy of the cascaded BS-IRS-user channels in these three schemes mainly depends on the  estimation of the IRS-user channels, A2-IRS-user channels, and IRS-user channels, respectively, in the channel estimation Phase II. Specifically, for both Scheme 1 and that in \cite{Dai}, the estimation of the IRS-user channels is based on the received pilot signals transmitted from the users to the BS via the IRS, while for Scheme 2, the estimation of the A2-IRS-user channels is based on the pilot signals transmitted from A2 to the users via the IRS. Since the signal power received at users is much larger than that at the BS due to the stronger channel power gain of the A2-IRS-user links as compared to that of the BS-IRS-user links, Scheme 2 is expected to achieve higher estimation accuracy than the other two schemes.

\section{Simulation Results}
\begin{figure}[t]
	\centering
	\includegraphics[width=3.7in]{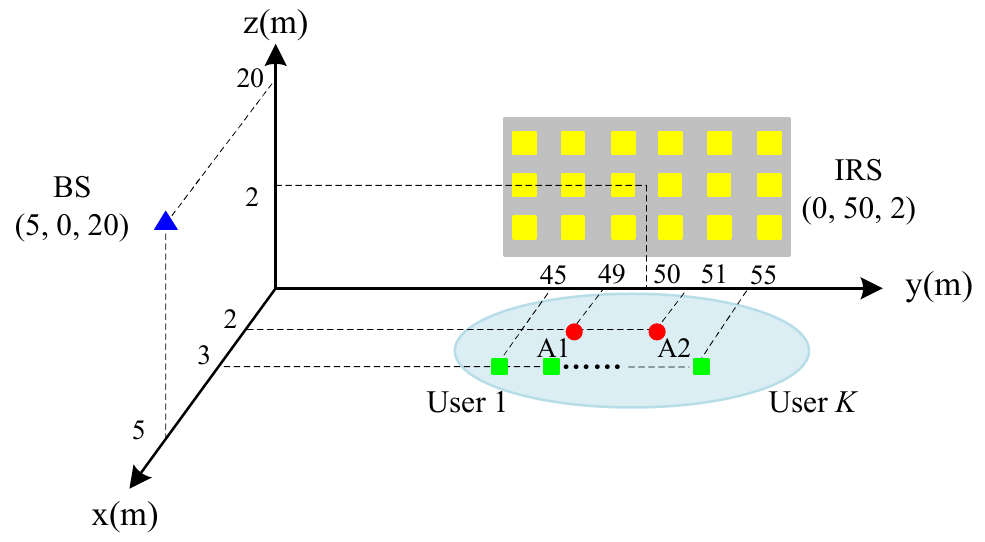}
	\caption{Simulation setup.}
	\label{simulation_setup}
	\vspace{5mm}
\end{figure}

In this section, we provide numerical results to evaluate the training overhead and channel estimation performance of the proposed two schemes. The simulation setup is shown in Fig. \ref{simulation_setup}. It is assumed that the BS, IRS (its central point), A1 and A2 are located at (5, 0, 20), (0, 50, 2),  (2, 49, 0) and (2, 51, 0) in meter (m), respectively, while $K$ users lie uniformly along the line from (3, 45, 0) m to (3, 55, 0) m. We assume that the IRS is equipped with a uniform planar array (UPA) composed of $8 \times 10$ reflecting elements; thus, we have $N=80$ (if not specified otherwise). The distance-dependent channel path loss is modeled as $\gamma=\gamma_0 d^{-\alpha}$, where ${\gamma_0}$ denotes the path loss at the reference distance of 1 m which is set as $\gamma_0=-30$ dB, $d$ denotes the individual link distance, and $c$ denotes the path loss exponent which is set as 2.5 for the links between the BS and A1/A2/users (due to the relatively large distance), set as 2.2 for the BS-IRS link (considering that the IRS is deployed vertically higher than A1/A2/users), and set as 2.1 for the other links (due to the short distance), by referring to the typical path loss exponent in \cite{WC2}. Assuming that the symbol rate for all transmission is $10^6$ symbols per second, the symbol duration is given by $10^{-3}$ ms. The coherence time for the channels associated with users is assumed to be $T_u=1$ ms, while that for the channels among the BS, IRS, A1 and A2 are assumed to be $T_c=500$ ms. The noise power at all receivers is set as $\sigma^2=-109$ dBm. Moreover, we consider the scheme proposed in \cite{Liu} and that proposed in\cite{Dai} (where the self-interference at the FD BS is assumed to be completely canceled) as benchmark schemes for performance comparison.

\begin{figure*}[t]
	\centering
	\vspace{0mm}	
	\subfigure[MTO versus $M$, with $(N,K)=(80,11)$.]{
		\includegraphics[width=1.9in]{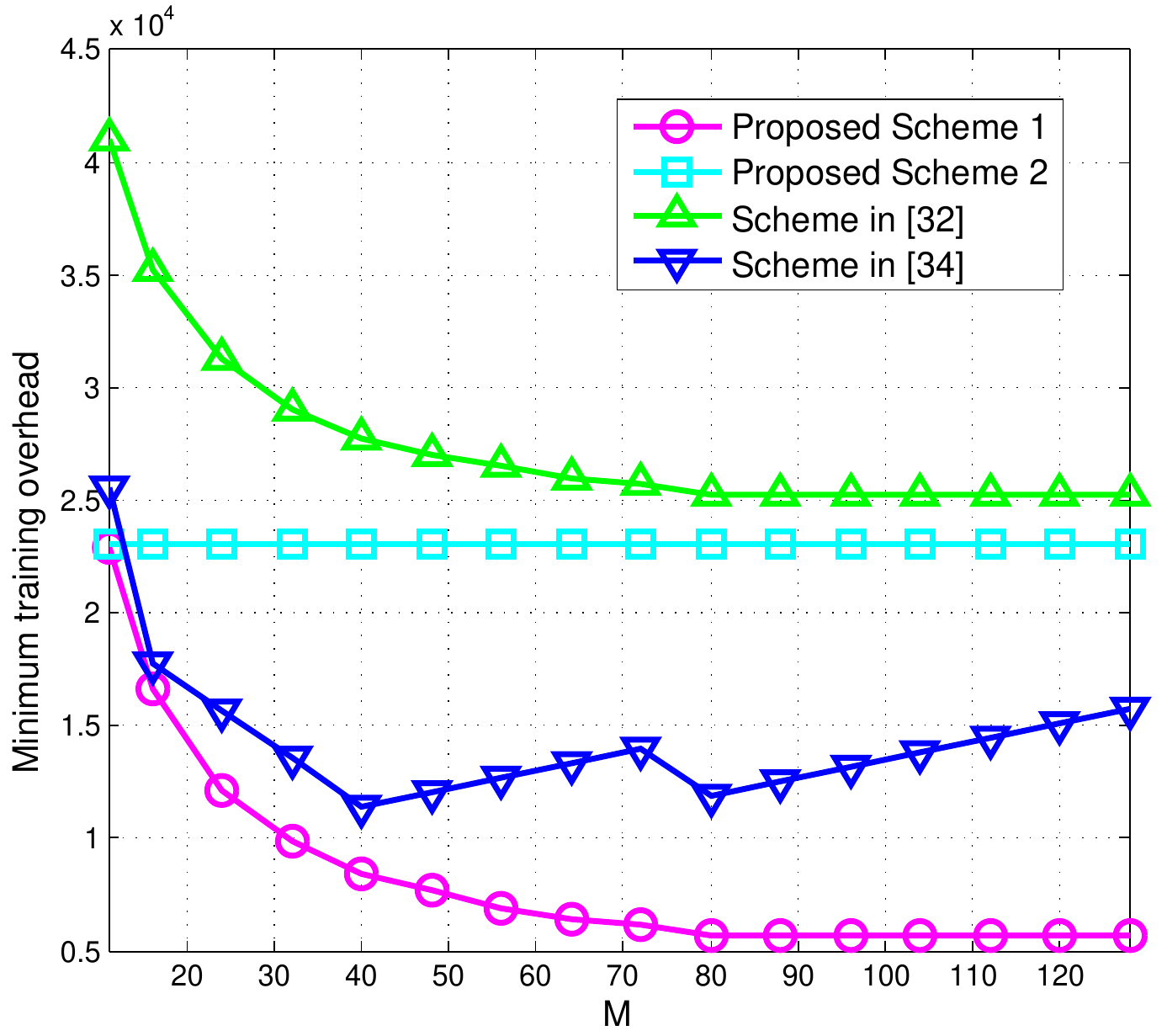}
		\label{fig:complexity_m}
	}	\hspace{1mm}
	\subfigure[MTO versus $N$, with $(M,K)=(80,31)$.]{
		\includegraphics[width=1.9in]{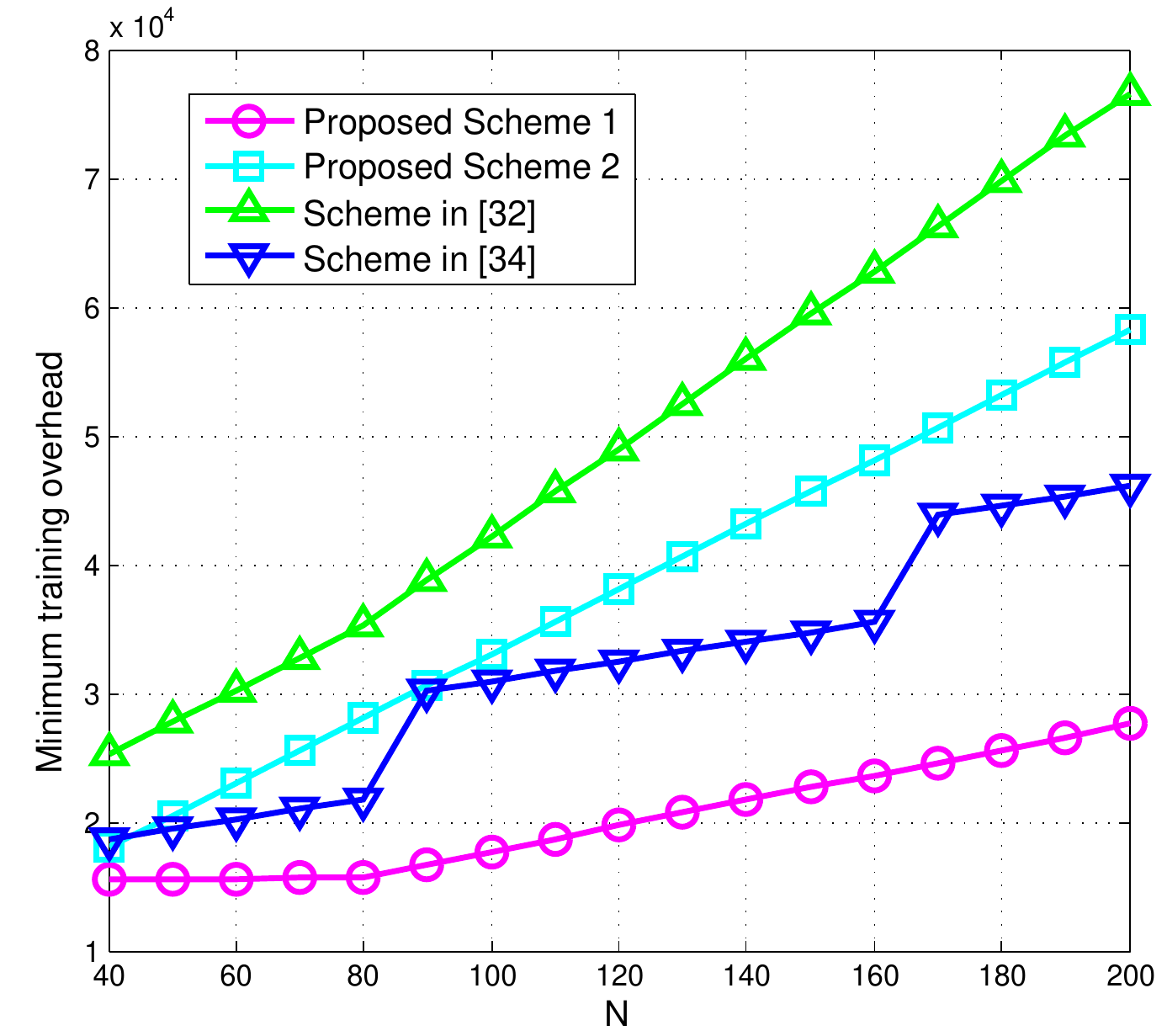}
		\label{fig:complexity_n}
	}\hspace{2mm}
	\subfigure[MTO versus $K$, with $(M,N)=(60,80)$.]{
		\includegraphics[width=1.9in]{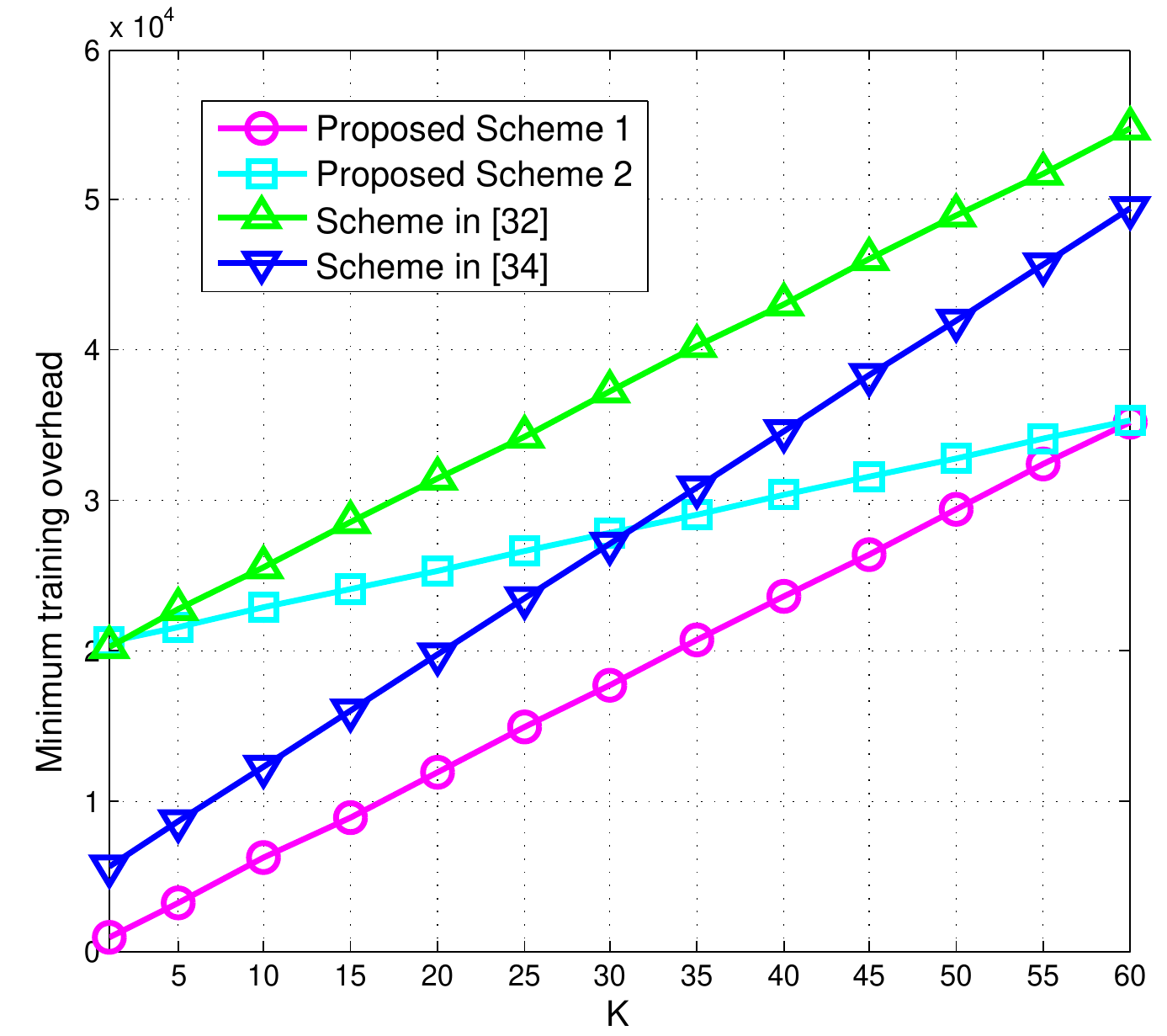}
		\label{fig:complexity_k}
	}
	\caption{Minimum training overhead (MTO) comparison between the proposed and benchmark schemes.}
	\label{complexity}
	\vspace{3mm}	
\end{figure*}

First, we compare the total training overhead (in terms of minimum number of pilot symbols in each $T_c$, i.e., those in Phase I and Phase II are both considered) for different schemes in Fig. \ref{complexity}. Specifically, Fig. \ref{fig:complexity_m} shows the required training overhead versus $M$, with $N=80$ and $K=11$. It is observed that both our proposed schemes (Schemes 1 and 2) significantly reduce the training overhead as compared to that of the scheme in \cite{Liu}. This is because the channel estimation Phase I is executed only once in the proposed schemes rather than periodically as in \cite{Liu}. Moreover, one can observe that Scheme 1 also outperforms the benchmark scheme in \cite{Dai}. This is because: (1) in Scheme 1, the training overhead for estimating the BS-IRS channel is $2(N+1)$, while that in \cite{Dai} is $M(N+Q)$; and (2) in Scheme 1, the IRS-user channels are jointly estimated via user-grouping and an orthogonal design of reflection coefficients and pilot symbols, which is different from that in \cite{Dai} where users transmit pilot symbols consecutively and the BS estimates each ${\bf h}_{su_k}$ individually. As such, multiple antennas at the BS are better exploited in Scheme 1 to reduce the pilot overhead. It is also observed that Scheme 2 has higher training overhead required than Scheme 1 under this setup, which is in accordance with Table I. Moreover, one can observe that the training overhead in \cite{Dai} does not monotonically vary with $M$. This is expected because in \cite{Dai}, as $M$ increases, the pilot symbols required in Phase I increases while that required in Phase II decreases, which thus leads to an optimal $M$ for achieving minimum training overhead.

\begin{figure*}[t]
	\centering	
	\subfigure[NMSE versus $p$, with $(M, N, K)=(128, 80, 11)$.] 
	{
		\includegraphics[width=2.8in]{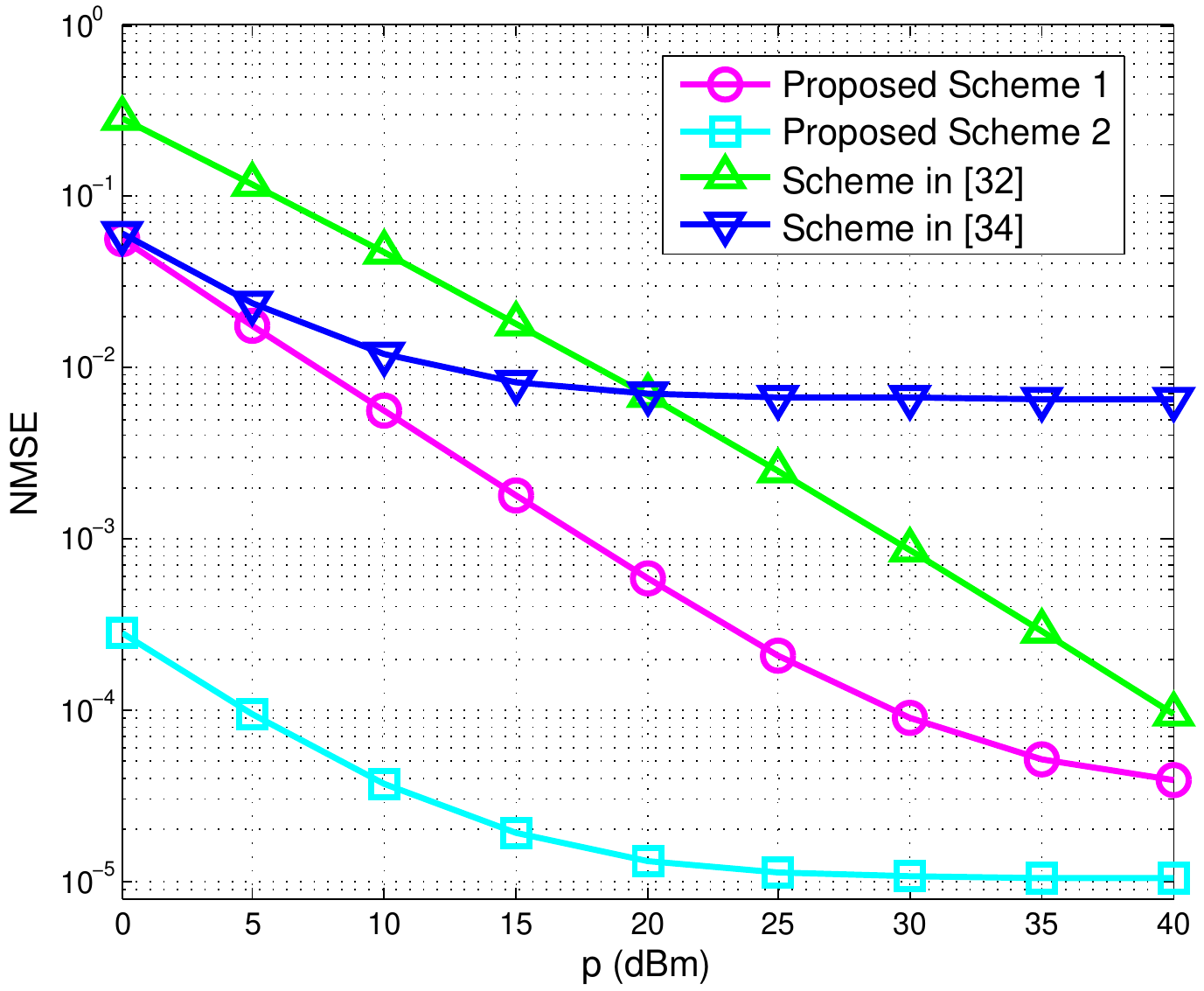}
		\label{fig:MSEVsP}
	}	\hspace{10mm}
	\subfigure[NMSE versus $M$, with $p=30$ dBm and $(N, K)\!=\!(80, 11)$.]
	{
		\includegraphics[width=2.8in]{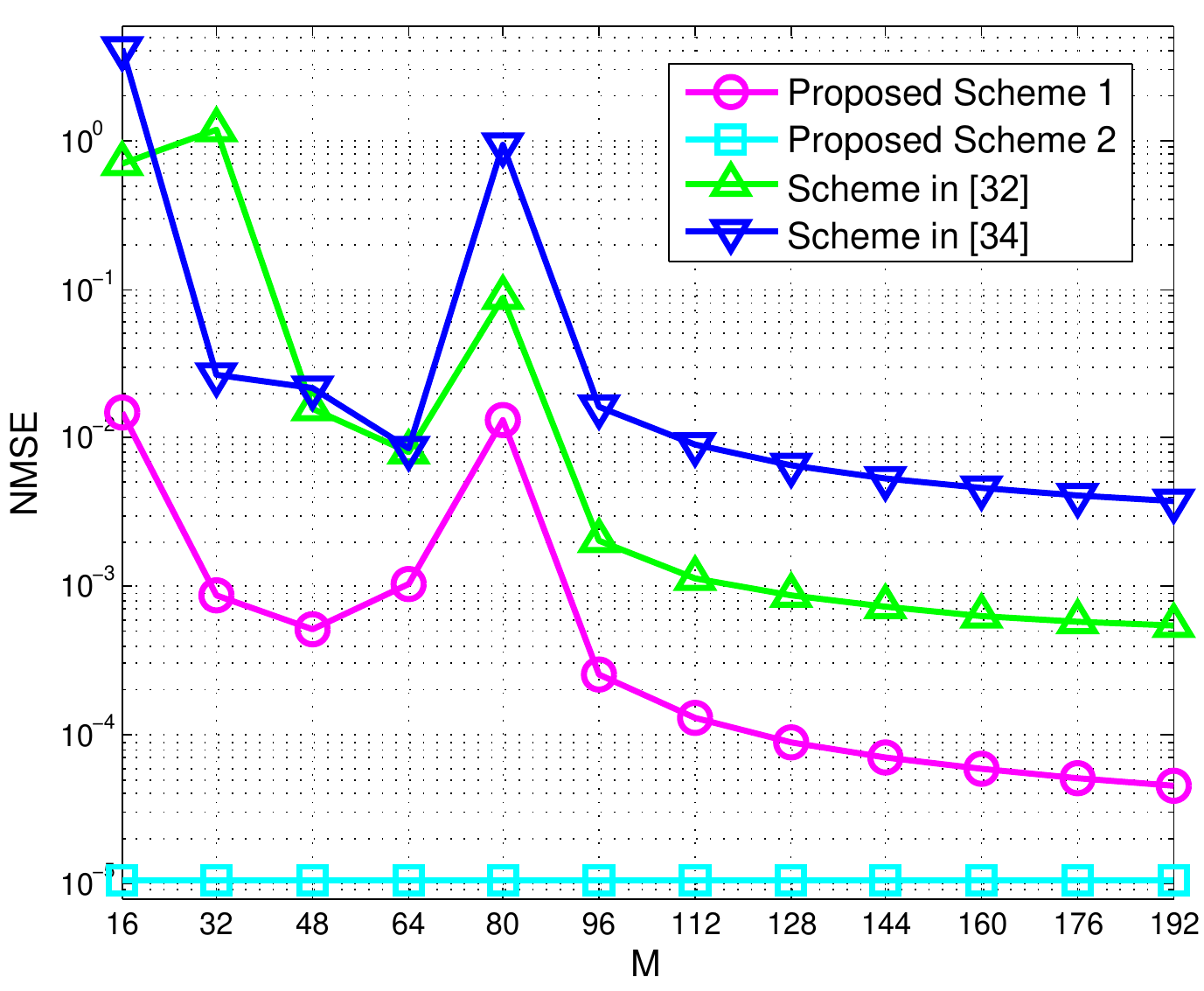}
		\label{fig:MSEVsM}
	}	\vspace{1mm}\\
	\subfigure[NMSE versus $N$, with $p\!=\!30$ dBm and $(M, K)\!=\!(128, 11)$.]
	{
		\includegraphics[width=2.8in]{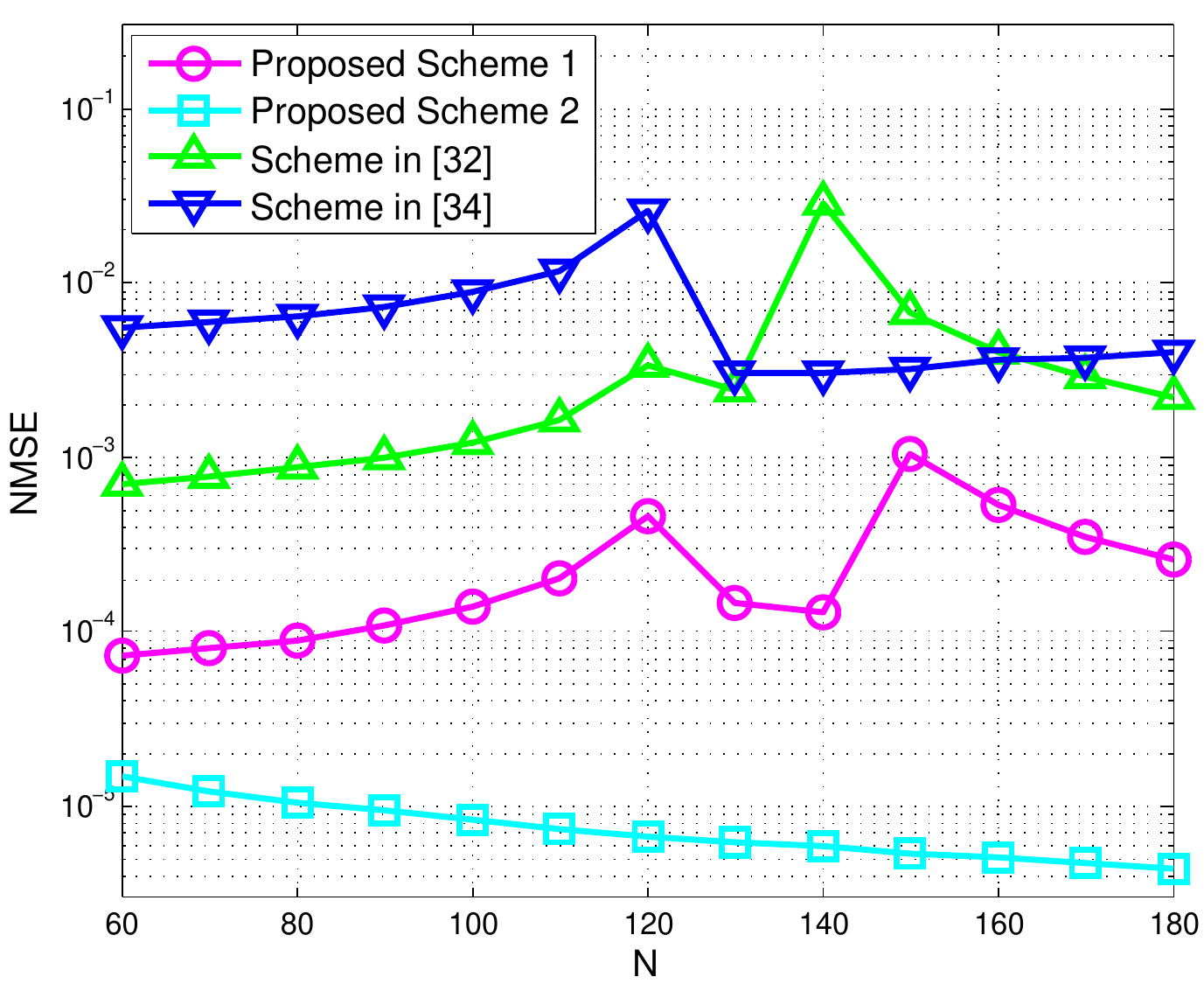}
		\label{fig:MSEVsN}
	}	\hspace{10mm}
	\subfigure[NMSE versus $K$, with $p\!=\!30$ dBm and $(M, N)\!=\!(128, 80)$.]
	{
		\includegraphics[width=2.8in]{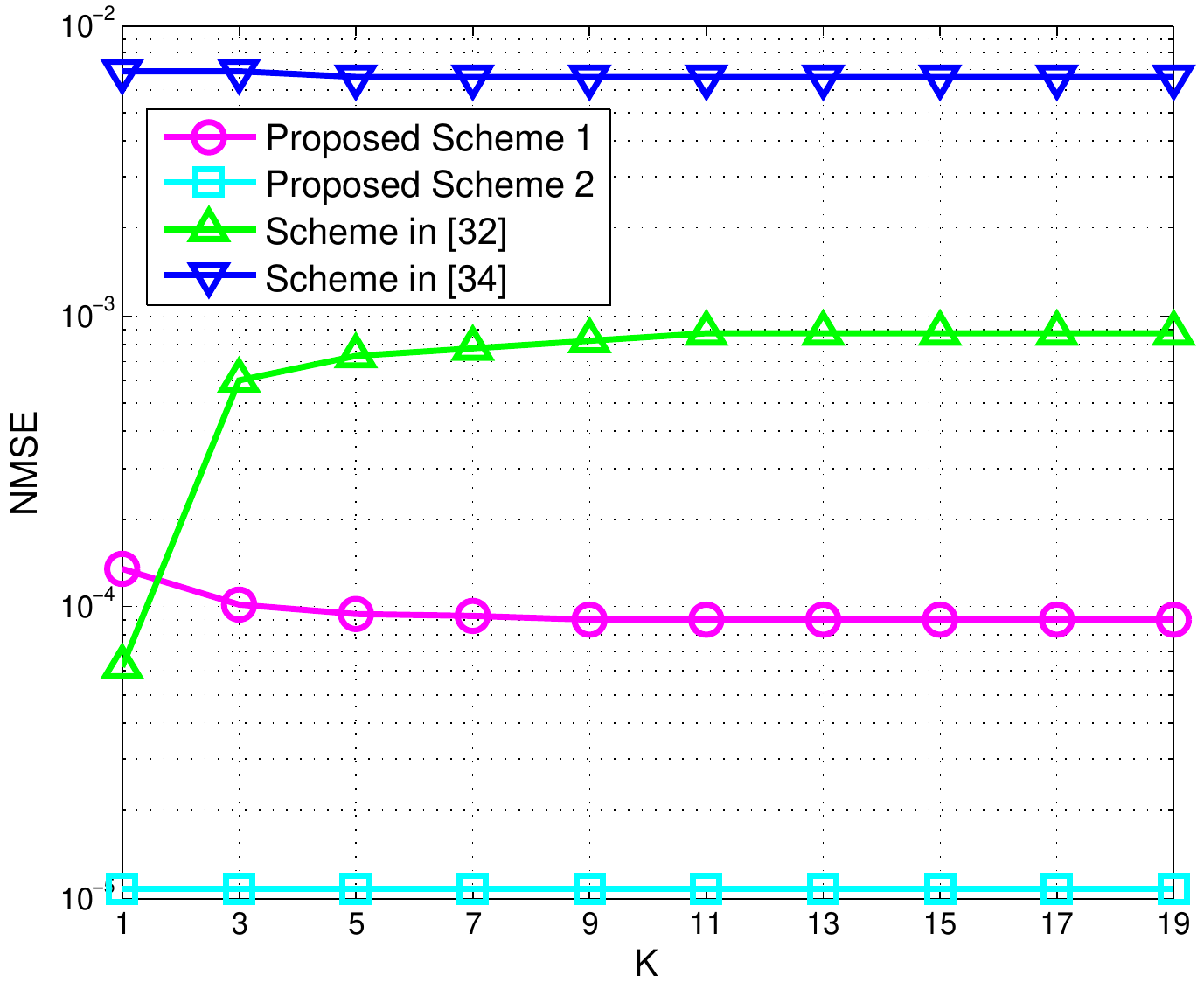}
		\label{fig:MSEVsK}
	}	\vspace{0mm}
	\caption{NMSE for cascaded BS-IRS-user channel estimation versus $p$, $M$, $N$ and $K$, respectively.}
	\label{MSE}
	\vspace{4mm}
\end{figure*}

Fig. \ref{fig:complexity_n} shows the training overhead  versus $N$, with $M=80$ and $K=31$. On one hand, it is observed that both our proposed schemes are more efficient than the scheme in \cite{Liu}. On the other hand, Scheme 1 outperforms the benchmark scheme in \cite{Dai} due to the similar reason given for Fig. \ref{fig:complexity_m}. Moreover, Scheme 1 is more efficient than Scheme 2 under this setup.  

Fig. \ref{fig:complexity_k} shows the training overhead versus $K$, with $M=60$ and $N=80$. First, it is observed that Scheme 1 still outperforms all the other schemes. Second, as $K$ increases, the training overhead of Scheme 2 increases more slowly than other schemes. This is because for the cascaded BS-IRS-user channel estimation in Scheme 2, only the anchor (i.e., A2) transmits pilot symbols and all users estimate the channels simultaneously, thus incurring a fixed pilot overhead as $N+1$, regardless of $K$; while only the training overhead for estimating the BS-user direct channels increases linearly with $K$. As a result, Scheme 2 becomes more efficient as the number of users increases. In particular, when $K\left\lceil\frac{N}{M}\right\rceil>N+1$ according to Table I, Scheme 2 even outperforms the benchmark scheme in \cite{Dai} and is on par with Scheme 1 when $K \geq M$.

Fig. \ref{MSE} shows the NMSE for the estimation of the cascaded channels ${\bf H}_{bsu_k}$'s versus $p$ (the pilot power channel estimation Phase II), $M$, $N$ and $K$, respectively, with the pilot power in channel estimation Phase I being setting as $p_1=p_2=40$ dBm. Note that each scheme is implemented with its minimum training overhead required as given in Table I. Moreover, for fair comparison, we assume that for the scheme in \cite{Dai}, the BS's pilot power for estimating the BS-IRS channel is also set as $40$ dBm. The NMSE is defined as\vspace{-0mm}
\begin{equation}
\text{NMSE}\!=\!\frac{\mathbb{E}\left\{\sum_{k=1}^K||{\bf H}_{bsu_k}\!-\!{\bf \hat H}_{bsu_k}||_F^2\right\}}{\mathbb{E}\left\{\sum_{k=1}^K||{\bf H}_{bsu_k}||_F^2\right\}}.\notag\vspace{0mm}
\end{equation}

In Fig. \ref{fig:MSEVsP}, it is observed that Scheme 2 outperforms Scheme 1, and as $p$ increases, the NMSE of both schemes first decreases and then tends to be saturated. The reason can be drawn from the computation of ${\bf{\hat H}}_{bsu_k}$ in these two schemes, i.e., (\ref{H_bsuk_est}) and (\ref{H_bsu_S2}), respectively. Specifically, in Scheme 1, the estimation error in the cascaded BS-IRS-user channels mainly depends on those in ${\bf \hat H}_{bs}$ and ${\bf \hat h}_{su_k}$ (note that the error in ${\bf \hat h}_{su_k}$ partially comes from that in ${\bf \hat h}_{bu_k}$ since the estimation of ${\bf  h}_{su_k}$ is based on the received signal at the BS by removing that from the direct BS-user channel using ${\bf \hat h}_{bu_k}$). As the pilot power for the  estimation of ${\bf h}_{bu_k}$ and  ${\bf h}_{su_k}$ increases, the NMSE in Scheme 1 is substantially reduced, until it is solely limited by the error in the estimated ${\bf \hat H}_{bs}$. On the contrary, for Scheme 2, (\ref{H_bsu_S2}) can be rewritten as $
{\bf{\hat H}}_{bsu_k}=\left({\bf{H}}_{bsa_1}+\Delta{\bf{H}}_{bsa_1}\right) {\rm{diag}}({\bf{ h}}_{a_2su_k}+\Delta{\bf{h}}_{a_2su_k})
({\rm{diag}}({\bf{ h}}_{a_1sa_2} + \Delta {\bf{ h}}_{a_1sa_2}))^{-1}$,
where $\Delta{\bf{H}}_{bsa_1}$, $\Delta{\bf{ h}}_{a_2su_k}$ and $\Delta{\bf{ h}}_{a_1sa_2}$ denote the estimation errors in ${\bf{\hat H}}_{bsa_1}$, ${\bf{\hat h}}_{a_2su_k}$ and ${\bf{\hat h}}_{a_1sa_2}$, respectively. As compared to Scheme 1, Scheme 2 has the following advantages in terms of NMSE: (1) the errors in the estimation of the direct BS-user channels do not propagate to that of the cascaded BS-IRS-user channels since they are independently estimated via training by users and A2, respectively; (2) the estimation of ${\bf{h}}_{a_2su_k}$ is more accurate than that of ${\bf{h}}_{su_k}$ in Scheme 1 since it is obtained based on pilot signals from A2 to users via the IRS instead of those from users to the BS via the IRS (the former channel is much stronger than the latter one due to the drastically shorter link distances); and (3) note that $\Delta{\bf{H}}_{bsa_1}$ and $\Delta{\bf{ h}}_{a_1sa_2}$ are determined by $p_1$ and $p_2$ in channel estimation Phase I, while $\Delta{\bf{ h}}_{a_2su_k}$ is determined by $p$ in channel estimation Phase II, thus when $p$ is relatively low such that $p \ll p_1$, we have $\Delta{\bf{ h}}_{a_1sa_2} \ll \Delta{\bf{ h}}_{a_2su_k}$. Considering the strong channel power gain of the A1-IRS-A2 link, the impact of errors $\Delta{\bf{H}}_{bsa_1}$ and $\Delta{\bf{ h}}_{a_2su_k}$ on ${\bf{\hat H}}_{bsu_k}$ is reduced due to the scaling over ${\bf{\hat h}}_{a_1sa_2}$. Owing to the above reasons, Scheme 2 achieves much lower NMSE than Scheme 1. However, similar to Scheme 1, the NMSE of Scheme 2 will be limited eventually by the errors in the estimated ${\bf{H}}_{bsa_1}$ and ${\bf{h}}_{a_1sa_2}$ as $p$ increases. 

On the other hand, both benchmark schemes have higher NMSE than our proposed schemes. In particular, for the scheme in \cite{Dai}, the distance between the BS and IRS is much longer than that between the IRS and anchors, thus the BS-IRS-BS dual-link estimation in \cite{Dai} is much less accurate than the anchor-IRS-BS/anchor estimation in our proposed schemes. For the scheme in \cite{Liu}, its NMSE performance depends on the estimation of the reference user's cascaded channel. Since the transmit power of the reference user is lower than that of the BS/anchors with fixed power supply, its channel estimation is less accurate than that in the proposed two schemes as well as that in \cite{Dai}. Thus, the scheme in \cite{Liu} performs the worst among the above discussed four schemes when $p$ is relatively low. However, as $p$ increases, it outperforms the scheme in \cite{Dai} and its NMSE can be substantially decreased due to the more accurate estimation of the reference user's cascaded channel.

The NMSE versus the number of antennas at the BS is plotted in Fig. \ref{fig:MSEVsM}, with $p=30$ dBm, $N=80$ and $K=11$. It is observed that both proposed schemes outperform the benchmark schemes. Specifically, when $M\ge N$, the NMSE for Scheme 1 and the two benchmark schemes all monotonically decrease with increasing $M$. The reason is that in the case of $M \ge N$, i.e., $M \ge 80$, all the three schemes estimate each cascaded BS-IRS-user channel individually based on one pilot symbol, thus there are $N=80$ unknowns in total to be estimated with $M$ observations obtained at the BS. Since $N=80$ observations are already sufficient for the estimation, as $M$ increases from 80, more observations are available which leads to $M-80$ redundant observations and thus substantial reduction in NMSE. On the other hand, when $M < N$, their required training overheads are shown in Table. I, where the numbers of additional observations do not monotonically vary as $M$ increases and thus result in the non-monotonic behavior of NMSE. However, for Scheme 2, increasing $M$ does not help decrease the NMSE since the computation of the cascaded channels relies on the estimation of the A2-IRS-user channels at users, which is independent of $M$ and thus leads to unchanged NMSE. Note that when $M$ becomes sufficiently large, Scheme 1 can even outperform Scheme 2 due to the above reason.

The NMSE versus the number of reflecting elements at the IRS is shown in Fig. \ref{fig:MSEVsN}, with $p=30$ dBm, $M=128$ and $K=11$. One can observe that the NMSE for Scheme 1 and the two benchmark schemes monotonically increases and non-monotonically varies with $N$ when $N\le M$ and $N>M$, respectively, due to the similar reason as that for Fig. \ref{fig:MSEVsM}. For Scheme 2, as $N$ increases, the training overheads in channel estimation Phase I and channel estimation Phase II both increase and thus the accuracy improves.

The NMSE versus the number of users is shown in Fig. \ref{fig:MSEVsK}, with $p=30$ dBm, $M=128$ and $N=80$. It is observed that both proposed schemes perform better than the benchmark schemes. In particular, as $K$ increases, the NMSE for  Scheme 1 decreases first and then tends to be saturated. This is because by increasing $K$, a larger number of pilot symbols are required for estimating all the BS-user channels and thus help improve the accuracy. Note that the cascaded BS-IRS-user channels are estimated based on the received signals at the BS by removing those from the direct channels using the estimated ${\bf h}_{bu_k}$'s, thus the improved estimation of ${\bf h}_{bu_k}$'s can help obtain more accurate estimation of ${\bf H}_{bsu_k}$'s and leads to lower NMSE. However, since the NMSE of Scheme 1 is also constrained by the errors in the estimated ${\bf H}_{bs}$ and ${\bf h}_{su_k}$'s, which are both independent of $K$ under this setup, the NMSE becomes saturated for sufficiently large $K$. For Scheme 2, the cascaded channels are computed based on the individually estimated ${\bf h}_{a_2su_k}$'s, which are insensitive to the number of users, thus the NMSE remains unchanged with respect to $K$. For the scheme in \cite{Liu}, as $N+1$ (which is much larger as compared to other schemes) pilot symbols are required to estimate the first reference user's channel, it outperforms the other three schemes when $K=1$. However, for the remaining $K-1$  users with $K>1$, the cascaded channels are estimated individually with reduced training overhead and thus increased error. As a result, as $K$ increases, the advantage of estimating the reference user's channel is averaged out and the NMSE increases until it is mainly constrained by the errors in the estimation of the $K-1$ (non-reference) users' channels. Finally, for the scheme in \cite{Dai}, since the channel estimation accuracy is dominated by the larger errors in the estimated ${\bf{H}}_{bs}$ due to long BS-IRS distance (as shown in Fig. \ref{fig:MSEVsP}), the improved estimation accuracy of the BS-user channels with increasing $K$ has no significant impact on the NMSE of the estimated cascaded channels. As a result, the scheme in \cite{Dai} exhibits the highest and almost unchanged NMSE under this setup among all considered schemes.

\begin{Rem}
	Note that all the four schemes considered do not directly estimate the cascaded BS-IRS-user channels for all users, but leverage the previously estimated BS-IRS-anchor/anchor-IRS-anchor/BS-IRS-BS channels or reference user's cascaded channel, which thus makes their NMSE performance affected by the accuracy of the previously estimated channels. Based on the observations in Fig. \ref{MSE}, it is shown that by exploiting the anchor-assisted training, the proposed two schemes outperform the benchmark schemes. The main reasons can be summarized as follows: (1) as compared to the estimation of the reference user's channel based on the pilots from the reference user (in \cite{Liu}), the estimation of BS-IRS-anchor and anchor-IRS-anchor channels (in our proposed schemes) can achieve higher accuracy because more transmit power is available at the anchors than that at the users and meanwhile a strong IRS-anchor channel can be established by deploying the anchors near the IRS; and (2) as compared to estimating the BS-IRS-BS channel for resolving the BS-IRS channel (in \cite{Dai}), estimating the BS-IRS-anchor and anchor-IRS-anchor channels (in our proposed two schemes) is more practically favorable, since the BS-IRS-anchor and anchor-IRS-anchor channels are much stronger than the BS-IRS-BS channel (due to the much shorter IRS-anchor distances). Moreover, it is also revealed that Scheme 1 benefits from increasing $M$ for not only the training overhead reduction, but also the NMSE performance improvement by exploiting the BS multi-antenna gains. Due to this reason, Scheme 1 can even achieve lower NMSE than Scheme 2 if $M$ is sufficiently large (e.g., massive MIMO BS).
\end{Rem}

\vspace{-0mm}
\section{Conclusion}
In this paper, we proposed two new anchor-assisted channel estimation schemes for IRS-aided multiuser communication systems. In the proposed Scheme 1, the common BS-IRS channel for all users is first estimated with sign uncertainty. Then the desired BS-IRS-user cascaded channels are resolved uniquely by leveraging this partial CSI. In contrast, the proposed Scheme 2 first estimates the BS-IRS-A1 and A1-IRS-A2 channels, and then estimates the A2-IRS-user channels for recovering the cascaded BS-IRS-user channels. The training overhead and feedback complexity were analyzed for both schemes, and their NMSE performances were evaluated by simulations, as compared to two recently proposed schemes in the literature. It was shown that Scheme 1 is most suitable when the BS has a large number of antennas, whereas Scheme 2 is more efficient otherwise. 

\section*{Appendix: Proof of Proposition 1 }
\vspace{-0mm}
There are in total $2^N$ possible estimations of ${\bf H}_{bs}$ since each ${{\bf{H}}_{bs}(1,n)}$ has two possible values. To show Proposition 1, we discuss the following two cases based on whether the actual ${\bf H}_{bs}$ equals to our constructed ${\bf \tilde H}_{bs}$ or not (with the noise ignored to focus on the issue of unique channel recovery).

{\bf Case 1:} ${\bf H}_{bs}={\bf \tilde H}_{bs}$. In this case, we have ${{\bf { H}}_{bs}(1,n)}=g_{1n}, \forall n$. Omitting the noise, we can rewrite the signal received at the BS in (\ref{y_est}) as \vspace{-1mm}
\begin{equation}
\label{case_a}
{\bf{\tilde y}}_b^{(i_4)}=\sqrt{p}{\bf H}_{bs}{\bf{h}}_{su_k} .\vspace{-1mm}
\end{equation}
Assuming that ${\bf \tilde H}_{bs}$ is full-rank, ${\bf{h}}_{su_k} $ is estimated as \vspace{-1mm}
\begin{equation}
\label{hat_h_suk}
{\bf{\tilde h}}_{su_k}=\frac{1}{\sqrt{p}}({\bf \tilde H}_{bs}^H{\bf \tilde H}_{bs})^{-1}{\bf \tilde H}_{bs}^H{\bf{\tilde y}}_b^{(k)}.
\end{equation}
Substituting (\ref{case_a}) into (\ref{hat_h_suk}), we have ${\bf{\tilde h}}_{su_k}={\bf{h}}_{su_k}$. Then, it is easy to verify that ${\bf{H}}_{bsu_k}={\bf{\tilde H}}_{bs}\text{diag}({\bf{\tilde h}}_{su_k})$ holds.

{\bf Case 2:} ${\bf H}_{bs} \ne {\bf \tilde H}_{bs}$. In this case, the actual ${\bf H}_{bs}$ is one of the other $2^N-1$ estimations except ${\bf \tilde H}_{bs}$. First, we define a set of vectors as $\{{\bf E}|[e_1,...,e_N] \in {\bf E}, e_n=\pm 1, \forall n, \text{and } [e_1,...,e_N] \ne {\bf 1}\}$. Then there are in total $2^N-1$ vectors in ${\bf E}$ and each of them is denoted by ${\bf e}_q$, $q=1,...,2^N-1$. As such, the other estimations of ${\bf H}_{bs}$ except ${\bf \tilde H}_{bs}$ can be expressed by
\begin{equation}
 \label{H_bs_bar}
 {\bf \bar H}_{bs}^{(q)}={\bf \tilde H}_{bs}\text{diag}({\bf e}_q), q=1,...,2^N-1.
\end{equation}
Here we provide an example to illustrate ${\bf \bar H}_{bs}^{(q)}$. Assume that  ${{\bf\bar{ H}}_{bs}^{(1)}(1,1)}=-g_{11}$ and ${{\bf\bar{ H}}_{bs}^{(1)}(1,n)}=g_{1n}$ for $\forall n \ne 1$, i.e.,
\begin{equation*}\small
\label{H_bs_3}
{\bf{\bar H}}_{bs}^{(1)}=\left[
\begin{aligned}
	&~-g_{11}         &g_{12}~~         &~~...&g_{1N}~~\\
	&-\alpha_{21}g_{11}&\alpha_{22}g_{12}&~~...&\alpha_{2N}g_{1N}\\
	&~~~~~~.            &.~~~~           &~~...&.~~~~~                \\
	&-\alpha_{M1}g_{11}&\alpha_{M2}g_{12}&~~...&\alpha_{MN}g_{1N}
\end{aligned}
\right].
\end{equation*}
Then, we have ${\bf{\bar H}}_{bs}^{(1)}={\bf{\tilde H}}_{bs}\text{diag}({\bf e}_1)$, where ${\bf e}_1=[-1,1,...,1]$.

Since ${\bf{H}}_{bs}$ must be one of the other ${\bf \bar H}_{bs}^{(q)}$'s, there must exists an ${\bf e}_q$ such that ${\bf{H}}_{bs}={\bf{\bar H}}_{bs}^{(q)}$. Without loss of generality, we denote ${\bf{H}}_{bs}={\bf{\bar H}}_{bs}^{(q^*)}$. Correspondingly, by omitting the noise, we can rewrite the BS received signal in (\ref{y_est}) as
\begin{equation}
  \label{y_b_4}
  {\bf{\tilde y}}_b^{(i_4)}=\sqrt{p}{\bf{\bar H}}_{bs}^{(q^*)}{\bf{h}}_{su_k}.
\end{equation}
Then, by substituting (\ref{y_b_4}) into (\ref{hat_h_suk}), the estimation of ${\bf h}_{su_k}$ is obtained as
\begin{equation}
{\bf{\tilde h}}_{su_k}=({\bf{ \tilde H}}_{bs}^H{\bf{ \tilde H}}_{bs})^{-1}{\bf{ \tilde H}}_{bs}^H{\bf{\bar H}}_{bs}^{(q^*)}{\bf{h}}_{su_k}.\label{hat_h_suk_2}
\end{equation}
Based on (\ref{H_bs_bar}) and (\ref{hat_h_suk_2}), ${\bf{\tilde h}}_{su_k}$ is rewritten as
\begin{equation*}
{\bf{\tilde h}}_{su_k}=({\bf \tilde H}_{bs}^H{\bf \tilde H}_{bs})^{-1}{\bf \tilde H}_{bs}^H{\bf \tilde H}_{bs}\text{diag}({\bf e}_{q^*}){\bf{h}}_{su_k}.\label{hat_h_suk_3}
\end{equation*}
Since $({\bf{ \tilde H}}_{bs}^H{\bf{ \tilde H}}_{bs})^{-1}{\bf{ \tilde H}}_{bs}^H{\bf{\tilde H}}_{bs}={\bf{I}}$, we have $
	{\bf{\tilde h}}_{su_k}=\text{diag}({\bf e}_{q^*}){\bf{ h}}_{su_k}$,
which implies that although ${\bf{\tilde h}}_{su_k}$ is not exactly the same as ${\bf{h}}_{su_k}$, the difference lies only in the sign of some/all of its elements. Based on the constructed ${\bf \tilde H}_{bs}$ and the estimated ${\bf{\tilde h}}_{su_k}$, the cascaded BS-IRS-user channel is resolved as\vspace{-0mm}
\begin{equation}
	{\bf{\tilde H}}_{bsu_k}={\bf \tilde H}_{bs}\text{diag}({\bf{\tilde h}}_{su_k})\mathop {\rm{ = }}\limits^{\rm{(a)}}{\bf \tilde H}_{bs}\text{diag}({\bf e}_{q^*})\text{diag}({\bf{h}}_{su_k}),\notag\vspace{-0mm}
\end{equation}
where (a) applies $\text{diag}(\text{diag}({\bf e}_{q^*}){\bf{h}}_{su_k})\!=\!\text{diag}({\bf e}_{q^*})\text{diag}({\bf{h}}_{su_k})$. Considering that ${\bf \tilde H}_{bs}\text{diag}({\bf e}_{q^*})={\bf{\bar H}}_{bs}^{(q^*)}={\bf{H}}_{bs}$, we have ${\bf \tilde H}_{bs}\text{diag}({\bf{\tilde h}}_{su_k})={\bf{H}}_{bs}\text{diag}({\bf{h}}_{su_k})={\bf{H}}_{bsu_k}$.

Based on Cases 1 and 2 in the above, it is concluded that ${\bf{H}}_{bsu_k}\!=\!{\bf \tilde H}_{bs}\text{diag}({\bf{\tilde h}}_{su_k})$ always holds, regardless of whether ${\bf{H}}_{bs}\!=\!{\bf{\tilde H}}_{bs}$ or not, which thus completes the proof of Proposition 1.

\ifCLASSOPTIONcaptionsoff
\newp\\a\apage
\fi

\end{document}